\documentclass[12pt,preprint]{aastex}
\shorttitle{H$_2$ Formation Pumping}
\shortauthors{Islam et al.}
\usepackage{rotating}
\begin{document}

\title{Formation Pumping of Molecular Hydrogen in Dark Clouds}

\author{Farahjabeen Islam}
\affil{University College London, Department of Physics and Astronomy, Gower 
Street, London WC1E 6BT}
\email{fislam@star.ucl.ac.uk}

\author{Cesare Cecchi\textendash Pestellini}
\affil{INAF ~\textendash~ Osservatorio Astronomico di Cagliari, Strada n.54, 
Loc. Poggio dei Pini, 09012 Capoterra (CA), Italy}
\email{ccp@ca.astro.it}

\author{Serena Viti}
\affil{University College London, Department of Physics and Astronomy, Gower
Street, London WC1E 6BT}
\email{sv@star.ucl.ac.uk}

\author{Silvia Casu}
\affil{INAF ~\textendash~ Osservatorio Astronomico di Cagliari, Strada n.54, 
Loc. Poggio dei Pini, 09012 Capoterra (CA), Italy}
\email{scasu@ca.astro.it}

\begin{abstract}
Many theoretical and laboratory studies predict H$_2$ to be formed in highly 
excited ro\textendash vibrational states. The consequent relaxation of excited
levels via a cascade of infrared transitions might be observable in emission
from suitable interstellar regions. In this work, we model H$_2$ formation 
pumping in standard dense clouds, taking into account the H/H$_2$ transition 
zone, through an accurate description of chemistry and radiative transfer. The
model includes recent laboratory data on H$_2$ formation, as well as the 
effects of the interstellar UV field, predicting the populations of 
gas\textendash phase H$_2$ molecules and their IR emission spectra. 
Calculations suggest that some vibrationally excited states of H$_2$ might be 
detectable towards lines of sight where significant destruction of H$_2$ 
occurs, such as X-ray sources, and provide a possible explanation as to why 
observational attempts resulted in no detections reported to date.
\end{abstract}


\section{Introduction}
By far the most abundant element in the universe is hydrogen. Consequently 
H$_2$ is the most abundant molecule and is the dominant collision partner in 
dark interstellar clouds. Dust grain surfaces act as heterogeneous catalysts in
the formation of H$_2$ molecules from atomic hydrogen \citep{GS63,HS70}. Beyond
this general consensus, the actual formation mechanism remains elusive and the 
internal energy distribution of the nascent hydrogen molecule is unknown. 
During formation, the H$_2$ binding energy $\Delta E_b = 4.476$~eV must be 
partitioned between the ro\textendash vibrational excitation and translational 
energy of the nascent molecule and heating of the dust grain \citep{DW93}. By 
studying the formation pumping of molecular hydrogen, namely the ro\textendash
vibrational distribution of nascent H$_2$ molecules, we can constrain 
interstellar chemistry both in the gas\textendash phase and on grain surfaces.

If the internal and translational energies of nascent molecules are relatively 
small, then significant grain heating must take place, which may lead to the 
desorption of volatile molecules from the dust grain surface \citep{DW93,R07}.
The H$_2$ internal energy distribution could have a significant impact on the 
chemistry occurring in the interstellar medium (ISM) because vibrationally 
excited H$_2$ will increase the overall energy budget of gas\textendash phase 
processes. There have been many theoretical and laboratory studies that 
predict H$_2$ to be formed in highly excited ro\textendash vibrational states 
(e.g., \citealt{S10,MT06} and references therein). It is possible that this 
formation pumping may be observable in the infrared (IR) spectra of H$_2$ 
molecules. 

The effects induced in the IR spectrum of H$_2$ by ro\textendash 
vibrational excitation of nascent molecules was first considered by 
\citet{BD76}, who employed a formation pumping model in which equipartition of 
the H\textendash H binding energy released was arbitrarily assumed. In this 
model, the binding energy is split equally between the internal energy of the 
molecule, its translational energy on desorption from the grain surface and the
heat imparted to the grain lattice. The molecule is formed at an effective 
temperature $T_f \sim 9000$~K. The internal energy is spread with a Boltzmann 
distribution throughout the ro\textendash vibrational levels, with the 
ortho\textendash to\textendash para ratio (OPR) being approximately 3. 
Subsequently, several classical molecular dynamics and quantum mechanical 
calculations have been carried out for H$_2$ formation on surfaces whose 
chemical compositions are analogous to interstellar dust grains, but the 
results have shown a wide dispersion in predicted vibrational distributions. 

\citet{DB96} proposed a ro\textendash vibrational distribution function that 
boosts the populations of vibrational states relative to the population of 
rotational states. The formation temperature is $T_f = 5 \times 10^4$~K, the 
OPR is $\sim 2.8$ and the mean 
vibrational and rotational levels are 5.3 and 8.7, respectively. \citet{DW86} 
suggested a mechanism which differs from previous proposals in the method of 
stabilization of the reacting complex. The stabilization energy ($\sim 0.4$~eV)
is transferred to a surface band, whose energy is that of the OH stretching 
vibration. The \citet{DW86} model predicts that the H$_2$ molecule on formation
is ejected into the gas vibrationally excited $(v \sim 6, 7)$ but rotationally 
cool $(J = 0, 1)$. The \citet{DW86} and \citet{BD76} models were employed by 
\citet{LB95} to generate H$_2$ IR spectra. \citet{LB95} presented results for 
another formation pumping scheme, which postulates that no energy is 
transferred to translation or to dust lattice modes. Thus, H$_2$ is formed in 
its highest vibrational level, $v = 14$, close to the dissociation threshold, 
with $J = 2$ and 3 weighted by the nuclear spin statistics. H$_2$ spectra were
also presented by \citet{T03} based on ER quantum calculations by \citet{F00} 
and \citet{M01}. Finally, \citet{TU01} constructed formation pumping models for
hydrogen molecules newly\textendash formed on icy mantles, carbonaceous and 
silicate dust, based on classical and quantum theoretical studies of molecular 
dynamics \citep{P98,T99,M01}. All these models provide characteristic spectral 
patterns, which may be used to discriminate between the H$_2$ formation pumping
mechanisms via astronomical observations.

\citet{DW93} suggested that the most favourable location to detect formation
pumping would be from dense, dark, quiescent, star\textendash less cores, 
where ultraviolet (UV) pumping is minimised. This suggestion is supported by 
\citet{TU01} and by \citet{T03}, who state that the relative emissivities of 
lines due to formation pumping in dense clouds can be a factor of 500 greater 
than in diffuse clouds. However, dark cloud observations have as yet failed to
detect any spectral lines due to formation pumping (e.g., \citealt{T03} and 
\citealt{C09}). In a laboratory study, \citet{C09} consider that 
non\textendash detections of H$_2$ in dark clouds may be explained by the 
thermalization of nascent H$_2$ molecules on the surface of dust grains, 
either within the porous structure of the grain or via collisions with other 
adsorbates in the thick icy mantle surrounding dust grains in dark clouds. 
Furthermore, spectra may be not observable due to the lack of an efficient 
mechanism to supply H atoms, such as a very high cosmic\textendash ray 
ionization rate $\zeta \sim 10^{-14}$~s$^{-1}$ to dissociate H$_2$ molecules, 
as found by \citet{LB95}. 

In this paper, we present a new experimentally derived formation pumping model 
constructed from the results of the UCL Cosmic Dust Experiment \citep{PD03,
C06} and we compute IR emission spectra of H$_2$ expected in 
interstellar clouds. The H$_2$ formation pumping model is coupled with an 
accurate description of both radiative transfer and chemistry in stratified 
dark clouds. Section 2 contains a description of the formation pumping 
excitation model employed in the calculations. In Section 3, we summarize the 
method and the procedures followed using different formation pumping models to
calculate the complete ro\textendash vibrational distribution of H$_2$ formed 
on dust grains. H$_2$ emission spectra are presented in Section 4. We discuss 
observational implications and we present our conclusions in Section~5.

\section{H$_2$ formation pumping}
\subsection{The experiment}
The ro\textendash vibrational excitation of molecular hydrogen desorbed from
surfaces can be measured experimentally. A quantitative partition of the 
excitation between vibrational states has been investigated by the UCL Cosmic 
Dust Experiment, which probes the distribution of the rotational states within
each vibrational manifold \citep{PD03,C06}. The experiment studies the 
formation of molecular hydrogen, primarily HD, on a highly\textendash oriented
pyrolitic graphite (HOPG) surface under ultrahigh vacuum, following continuous 
irradiation of the surface by H and D atoms. The nascent HD (or H$_2$) molecule
will desorb in ro\textendash vibrational states $(v, J)$ of the ground 
electronic state. Hydrogen molecules are state\textendash selectively ionised 
using laser\textendash induced resonance enhanced multi\textendash photon 
ionisation spectroscopy. The relative populations of the ro\textendash 
vibrational states are then derived from ion yields.

The experiment probes HD preferentially to H$_2$ (or D$_2$) because there is a 
significant amount of undissociated H$_2$ originating from the H\textendash 
atom source. As studies of $v = 1$ and 2 for the isotopic species revealed 
very similar flux densities and rotational distributions for both HD and H$_2$ 
molecules \citep{C06}, we extrapolate the HD $v = 3 - 7$ data to obtain the 
ro\textendash vibrational distribution of nascent H$_2$. However, nascent HD in
the $v = 0$ state could not be detected above the signal from background gas in
the vacuum chamber (see \citealt{C06}). As a modest estimate of the internal 
excitation of H$_2$, we set the $v = 0$ populations to be equal to the 
populations of the $v = 1$ states. Since this is an arbitrary choice, H$_2$ 
emission spectra have been generated with the $v = 0$ states both significantly
more and less populated than expected. The sensitivity of the H$_2$ spectra to 
the error in the population of the $v = 0$ states is discussed further in 
Section~4.

HD formed on HOPG held at 15~K has been found in ro\textendash vibrational 
states $(v, J) = (1 - 2, 0 - 4)$, $(3 - 5, 0 - 6)$, $(6, 0 - 4)$ and $(7, 0-3)$
\citep{C06,I07,L08}. The ro\textendash vibrational distribution of HD peaks at 
$v=4, \, J=1$. The total vibrational population, found by summing the relative 
rotational populations, approximately doubles with subsequent $v-$state up to 
$v = 4$, then falling sharply at $v = 5$. This indicates that the vibrational 
distribution differs dramatically from a Boltzmann distribution. Examination of
Boltzmann plots reveals that the rotational distributions within each 
$v-$state also slightly deviate from a Boltzmann distribution. Translational 
energy of the molecule is known to have an upper limit of 0.9 eV \citep{C06}. 
Such a low translational energy of nascent hydrogen molecules has been 
observed by other experimental studies (\citealt{V04}). From assuming that the
$v = 0$ populations are roughly equal to the $v = 1$ populations, the 
ro\textendash vibrational distribution corresponds to an internal excitation 
of 1.74 eV, with at least 41\% of the HD binding energy flowing into the 
surface. 

In order to retain any deviation from the Boltzmann distribution, we scale the 
HD rotational populations by the appropriate nuclear spin statistical weight 
$g_N$ to obtain the rotational populations of H$_2$. Assuming the 
OPR of H$_2$ formation in space to be 
approximately 3 (e.g., \citealt{DB96} and \citealt{T01}), we set $g_N = 1$ and 
3 for even$-J$ and odd$-J$, respectively. By scaling the ro\textendash 
vibrational populations from HD to H$_2$, we note that although the 
ro\textendash vibrational distribution is preserved, the average energy of the 
molecule has marginally increased. This change in internal energy is due to the
energy levels of H$_2$ being more spaced out within the potential well, as 
H$_2$ is lighter than HD. Thus the resulting average energy of H$_2$ is 
1.95~eV. Three body coupling between 
the two H atoms and a quantum defect within the substrate may be strong 
enough to randomize the spin orientation of the newly formed molecule. However,
we do not expect there to be as many defects on our HOPG surface as on a real 
interstellar dust grain (see Section 2.2). Experiments conducted probing the 
ro\textendash vibrational distribution of H$_2$ as well as HD \citep{C06} show
that there is no significant difference in vibrational level populations for 
H$_2$ and HD formed in $v = 1$ and 2. Also, rotational temperatures of H$_2$ 
and HD within each vibrational state were found to be similar ($\sim 300$~K), 
taking an OPR equal to 3, for H$_2$. Furthermore, \citet{TU01} calculate that 
the OPR of newly\textendash formed H$_2$ is nearly 3 on silicate, carbonaceous
and icy surfaces. Thus, we feel that statistical weighting is a good 
approximation. 

\begin{figure}
\epsscale{.80}
\plotone{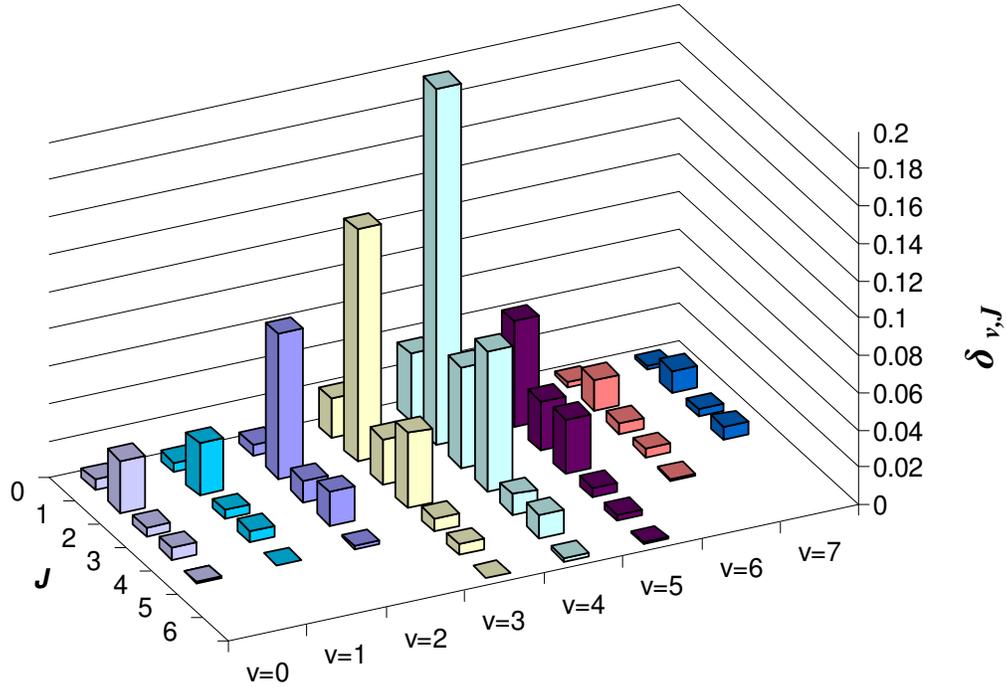}
\caption{The new formation pumping model. The relative ro\textendash 
vibrational populations for H$_2$ are extrapolated from experimental studies of
HD (see text) by assuming an ortho$-$to$-$para ratio of 3 and choosing
$\delta_{0,J} = \delta_{1,J}$.}
\label{fig1}
\end{figure}

The relative populations of all the states are then normalised to obtain 
$\delta_{vJ}$, the fraction of H$_2$ formed on grain surfaces that leaves the 
grain in level $(v,J)$, such that $\sum_{v,J} \delta_{vJ} = 1$. The formation 
pumping population distribution of H$_2$ used throughout this paper is shown in
Fig. \ref{fig1} and reported in Table~1.

\subsection{Astrophysical surfaces} 
Astrophysical surfaces are of course likely to be very different from the HOPG
surface used in the UCL Cosmic Dust Experiment. Important characteristics of 
an interstellar dust grain are chemical composition, fraction of 
crystallization, roughness, state of charge and temperature. Therefore, it is 
important to discuss how the adopted representation of the H$_2$ formation 
pumping population distribution may be related to the properties of 
interstellar grain surfaces. 

The experimental results utilised in this work \citep{C06,I07,L08} show that 
HD forms vibrationally excited on HOPG, which is a non\textendash porous, 
well\textendash ordered crystalline structure. \citet{A07} have shown that 
D$_2$ forms vibrationally excited $(v \ge 2)$ on non\textendash porous 
amorphous ice (NP ASW) held at $8 - 30$~K. Therefore, vibrational excitation of
nascent molecules seems to occur regardless of crystallization fraction. Both 
the UCL Cosmic Dust Experiment and \citet{A07} have detected nascent molecular 
hydrogen in vibrational states $v = 1 - 7$. The detection of 
vibrationally excited molecules in the two experiments implies 
that the formation process on HOPG and NP ASW may be similar. Consequently, 
internal excitation of nascent hydrogen molecules may not significantly 
depend on the chemical composition or degree of order of the surface, given a 
non\textendash porous surface. Unfortunately there is no work to date 
which quantifies the ro\textendash vibrational distribution of nascent 
molecular hydrogen on astrophysically relevant surfaces, other than the UCL
Cosmic Dust Experiment. \citet{G96} have shown that H$_2$ forms ro\textendash
vibrationally excited on 
carbon surfaces with temperatures of $90 - 300$~K, but these temperatures are 
too high to simulate the majority of interstellar conditions. \citet{V06} have 
found that the recombination of hydrogen is efficient on carbon surfaces at 
temperatures $11 - 18$~K, on olivine surfaces at temperatures $6 - 9$~K, on low
density amorphous ice surfaces at temperatures $11 - 15$~K and on high density 
amorphous ice surfaces at temperatures $14 - 18$~K. However, the
temperature\textendash programmed desorption (TPD) experiments of \citet{V06} 
do not probe the ro\textendash vibrational distribution of nascent molecules. 

The roughness of a dust grain may also affect the ro\textendash vibrational 
distribution of nascent molecules, as defects in the crystalline structure may
allow atoms to chemisorb, hence bond more strongly with the graphite surface. 
There is a 0.2 eV barrier to chemisorption on graphite surfaces \citep{Z02}, 
which arises from carbon atoms having to pucker out of the graphite sheet to 
bond with incident hydrogen atoms. Hence for the UCL Cosmic Dust Experiment, 
where the incident H\textendash~ and D\textendash atoms are at $T \sim 300$~K,
chemisorption is improbable unless the formation mechanism is dominated by 
reactions at defects. TPD of an etched graphite surface irradiated by 
H\textendash atoms at 2000~K was conducted by \citet{Z02}. The high 
temperature of the H\textendash atoms allowed them to overcome the 0.2 eV 
barrier and chemisorb to the surface. The etched surface created terrace edges 
on the graphite lattice, namely defects. \citet{Z02} found that atoms bond more
strongly at the defects. However, there was minimal change in the recombination
of chemisorbed atoms with the density of terrace edges on the graphite surface,
implying that the formation process was dominated by conventional sites on the 
planar surface. Hydrogen atoms adsorbed at terrace edges were found to desorb 
primarily in the form of hydrocarbons, rather than molecular hydrogen. However,
in dark clouds, the energy of the incident atoms is much lower than in the 
\citet{Z02} experiment. Therefore, it is still feasible that defects dominate 
the reaction of hydrogen atoms at low temperatures. In this case, theoretical 
work based on chemisorption and the Eley\textendash Rideal mechanism, such as 
that by \citet{TU01}, may be more accurate than results from the UCL Cosmic 
Dust Experiment. 

The charge state of a grain is likely to affect surface chemistry (e.g., 
\citealt{CH10}). If the charge on the grain is largely delocalized, then one 
would not expect there to be much effect on the recombination of two 
H\textendash atoms. If the charge is localized near the H$_2$ formation site, 
then it may be harder for the molecule to escape the grain, hence reducing the 
formation rate.

A more important factor is the dust grain morphology, in particular the 
porosity of the surface. Both bare and icy grains may exhibit a microporous 
structure \citep{Gr02,WH02,Wi07}. Although atoms are mobile on porous grains 
\citep{M08} and recombination is efficient \citep{H03}, the nascent hydrogen 
molecules are found to thermalize in the pores, losing kinetic and internal 
energy. Therefore, although the H$_2$ molecule may originally form with 
ro\textendash vibrational excitation, subsequent collisions with pore walls 
may mean that there is no apparent formation pumping of molecules that have 
escaped the grain surface. \citet{C09} have shown experimentally that D$_2$ 
does not form vibrationally excited on porous surfaces. 

The work presented in this paper incorporates our new formation pumping model
from the UCL Cosmic Dust Experiment, as well as some theoretical models, such 
as the work put forward by \citet{TU01}. We also include a model in which 
there is no apparent formation pumping. We compare spectra generated by these 
different formation pumping models in order to find observational markers to 
discriminate between formation pumping mechanisms. 

\section{The H$_2$ level distribution model}\label{models}
We have constructed models to compute the H$_2$ level distribution expected in
a thermally excited gas, based on a radiative transfer code developed to study 
H$_2$ formation pumping \citep{C05} and excitation in turbulent diffuse
interstellar clouds \citep{CCP05}. The approach is similar in some respects to 
those developed by \citet{SD89} and \citet{DB96}. However, we do not adopt
the cascade efficiency factor formalism \citep{BD76} and we solve the full
set of statistical equilibrium equations for the first ${\cal N} = 300$ levels 
of H$_2$. We describe the depth\textendash dependent H$_2$ photodissociation 
rates by self\textendash shielding functions \citep{vDB88,SD95}, including the 
prescription for line overlap given in \citet{DB96}. All radiation induced 
processes have been computed taking into account dust extinction, which has 
been assumed to follow the mean galactic interstellar extinction curve (e.g., 
\citealt{FM07}).

In statistical equilibrium the ${\cal N}$ populations $n_{i}$, $i=(v,J)$, of 
the H$_2$ levels are solutions to the set of algebraic equations
$$
n_i \{ \sum_{j < i} \left( A_{ij} + C_{ij} + W_{ij} \right) + \sum_{j > i}
\left( C_{ij} + W_{ij} \right) 
+ \beta_i + \zeta + {\cal D}_i + {\cal K}^-_i \} = 
$$
\begin{equation}
= \sum_{j > i} n_j A_{ji} + \sum_{j \ne i} n_j \left(C_{ji} + W_{ji} \right)+ 
{\cal K}^+_i + {\cal R} n_{\rm H} n_1 \delta_i 
\label{SEE}
\end{equation}
In eq.(\ref{SEE}), $W_{ij}$ are the excitation rates from the level $i$ to
level $j$ via UV pumping to electronically excited states, $A_{ij}$
are the Einstein coefficients for spontaneous radiative decay, $C_{ij}$ are the
temperature\textendash dependent collisional rates, $\beta_i$ is the rate of
photodissociation out of the level $i$, ${\cal D}_i$ is the rate of
additional destruction processes, such as collisional dissociation and
ionization, and ${\cal K}^\pm_i$ are entry (+) and exit (-) chemical rates. 
The cosmic\textendash ray destruction rate is denoted by $\zeta$.
The last term in the r.h.s. of eq.(\ref{SEE}) describes the formation of H$_2$ 
via grain catalysis:  ${\cal R}$ is the formation rate, $n_{\rm H}$ the total 
volume density of hydrogen, $n_1$ the volume density of atomic hydrogen, and 
$\delta_i$ is the fraction of H$_2$ formed on grain surfaces that leaves the 
grain in level $i$. The highest lying state is $(v,J) = 
(3,27)$ at about 52,000~K above the ground state. The level populations are 
subject to the normalization conditions $\sum_{vJ} n_{vJ} = n_{\rm H_2}$, 
where $n_{\rm H_2}$ is the number density of hydrogen molecules. 

The radiative transfer code includes inelastic collisions with H, He and 
ortho\textendash~ and para\textendash H$_2$, with fully quantum mechanical 
calculations of collisional rates given in studies by \citet{F98}, 
\citet{FR98}, \citet{FR98b}, \citet{FRZ98} and \citet{LB99}. For levels where 
quantum calculations are not available, the extrapolation scheme for the 
H\textendash H$_2$ rate collisions provided by Le~Bourlot and co\textendash 
workers in their code for photon$-$dominated regions (PDRs) has been adopted 
(Le~Bourlot private communication). For all other collisional partners when 
quantum calculations are lacking, the collision scheme put forward by 
\citet{T97} has been employed. For models with temperatures larger than 30~K, 
we have incorporated three rate coefficients for ortho\textendash para 
conversions provided by \citet{SD94}. Energy levels, radiative decay rates and 
dissociation probabilities for electronic transitions have been published by 
\citet{A92}, \citet{A93}, \citet{A93b} and \citet{A00}. Extra data, covering 
levels up to $J = 25$, were kindly provided by Abgrall (private 
communication). Quadrupole radiative decays and energies of ro\textendash 
vibrational levels of the ground electronic state were taken from 
\citet{WSD98}.

The H$_2$ formation rate is more fully described by moment equations 
\citep{LP09}, rather than by rate equations using the expression 
${\cal R} n_{\rm H} n_1 \delta_i$. This is because small grain sizes and low 
atom fluxes are subject to large fluctuations, and thus calculating the H$_2$ 
formation rate requires stochastic methods. However, \citet{LP09} find that the
moment equation results agree with the rate equation results in a wide range of
conditions, except for dust grains at temperatures larger than 18~K, in which 
case the rate equations overestimate the H$_2$ formation rate. For dark 
clouds, as investigated in this paper, where dust temperature are $\sim 10$~K
throughout the cloud, the H$_2$ formation rate can be adequately described 
using the standard rate equation term assumed in eq.(\ref{SEE}). Of course, if
the radiation environment differs substantially from the standard interstellar 
conditions, dust temperatures may be noticebly larger than 
20~K (see Section 4).

In the present study, for all formation pumping models, except the one 
presented in Section 2, the relative (normalized) populations of the 
ro\textendash vibrational states are given by 
\begin{equation}
\delta_{vJ} = {\cal C} g_N f_1(v,J)  \, {\rm exp}[-f_2
(\Delta E_{vJ}, T_f)]
\label{lev}
\end{equation} 
where ${\cal C}$ is a normalization constant and
$\Delta E_{vJ}$ is the energy in K of level $(v,J)$ referred to the ground
state. $f_1$ and $f_2$ are shape functions depending on the specific formation 
model. We consider the following H$_2$ formation pumping models:

\begin{itemize}

\item[$(i)$]
the formation pumping model described in Section 2, where $\delta_{vJ}$ is 
taken directly from extrapolation of HD experiments (see Table~1);

\item[$(ii)$]
the acquired internal energy, $E_i$ is statistically distributed among the 
energy levels, $f_1 = 2J + 1$ and $f_2 = \Delta E_{vJ}/T_f$; by setting the 
formation temperature to $T_f$ = 9,000~K the model proposed by \citet{BD76} is
recovered
\begin{equation}
\frac {\sum_{vJ} g_N \times \left(2J+1 \right) \Delta E_{vJ}\, 
{\rm exp} \left(-\Delta E_{vJ}/T_f \right)}
{\sum_{vJ} g_N \times \left(2J+1\right) \, {\rm exp} \left(-\Delta E_{vJ}/T_f 
\right)} \sim \frac {1.5 \, {\rm eV}}{\kappa_B},
\end{equation}
$\kappa_B$ being the Boltzmann constant;

\item[$(iii)$]
to enhance the populations of high $v$ states with respect to high $J$ states, 
we set $f_1 = v + 1$; this class of formation models with $T_f = 50,000$~K 
provides the pumping profile suggested by \citet{DB96};

\item[$(iv)$]
the three ro\textendash vibrational population distributions for H$_2$ 
newly\textendash formed on carbonaceous dust $(iv-c)$, silicate dust $(iv-s)$ 
and icy mantles $(iv-i)$ given by \citet{TU01}, model A; the functions $f_1$ 
and $f_2$ are taken from \citet{TU01};
 
\item[$(v)$]
the same as in $(iv)$ but for \citet{TU01} model B;

\item[$(vi)$]
a minimal H$_2$ formation pumping model which limits the ro\textendash
vibrational distribution to the lowest possible levels $(v,J)=(0,0)$ and (0,1) 
in the ratio 1:3; $f_1 = f_2 = 0$ except for $(v,J) = (0,0)$ and $(0,1)$ when 
$f_1 = 1$; this represents the ``no formation pumping" case. If nascent 
H$_2$ molecules in dark clouds thermalize with the dust grain pores or the icy 
mantle, as put forward \citet{C09}, then no formation pumping would be 
detected.
 
\end{itemize}

\section{H$_2$ IR emission from dark clouds}
The cloud is assumed to be a plane\textendash parallel slab of constant 
density, $n_{\rm H}$. The cloud is two\textendash sided illuminated by 
an isotropically incident UV radiation field. The field intensity is 
assumed to be the UV field scaling parameter, $\chi$, times the \citet{D78} 
estimate of the mean interstellar radiation field as reported in \citet{SD89}. 
The attenuation of the field due to dust extinction is computed using the 
analytical solution to the transport equation in plane\textendash parallel 
geometry given in \citet{FRR80}. The H/H$_2$ ratio calculations as a function 
of depth have been supplemented with the time\textendash~ and depth\textendash 
dependent gas\textendash grain UCL\_CHEM chemical model \citep{Vi04}. In
addition, we use the UCL\_PDR code \citep{Bell06} in order to get the cloud 
thermal profile. We find that gas temperatures are higher than $\sim 10$~K 
only at the very edge of dark clouds. However, in the case of translucent 
clouds, or non\textendash standard illumination ($\chi > 1$), gas temperatures
may be significantly larger than 10~K throughout the cloud. 

We perform calculations for translucent and dark clouds of number densities 
$n_{\rm H} = 10^3 - 10^6$~cm$^{-3}$, cosmic\textendash ray ionization rate 
$\zeta = 3 \times 10^{-17}$~s$^{-1}$ and column densities $N_{\rm H} = 4.8 
\times 10^{21} - 1.6 \times 10^{23}$~cm$^{-2}$, corresponding to visual 
integrated magnitudes $A_V = 3 - 100$ (e.g., \citealt{S06}). Dark clouds were 
chosen following the suggestion by \citet{DW93} and \citet{T03} that 
quiescent, dense clouds with no UV pumping are the regions of the ISM best 
suited to detect formation pumping. We extend the calculations to clouds with 
moderate extinctions in order to get larger formation pumping rates, although 
contamination from radiative pumping may be substantial (as is actually the 
case). For such clouds, the UCL\_PDR code in the density range $10^3 - 
10^6$~cm$^{-3}$ provides thermal profiles in which the ``10~K edge" is reached
at $A_V = 1 -2.5$~mag, measured from the external boundary of the cloud.
For $\chi \ga 100$, the UCL\_PDR code also finds that dust temperatures are 
larger than 20 K. In that case the use of rate equations for H$_2$ formation 
may be not accurate \citep{LP09}.

\begin{figure}
\begin{tabular}{cc}
\includegraphics[scale=0.43]{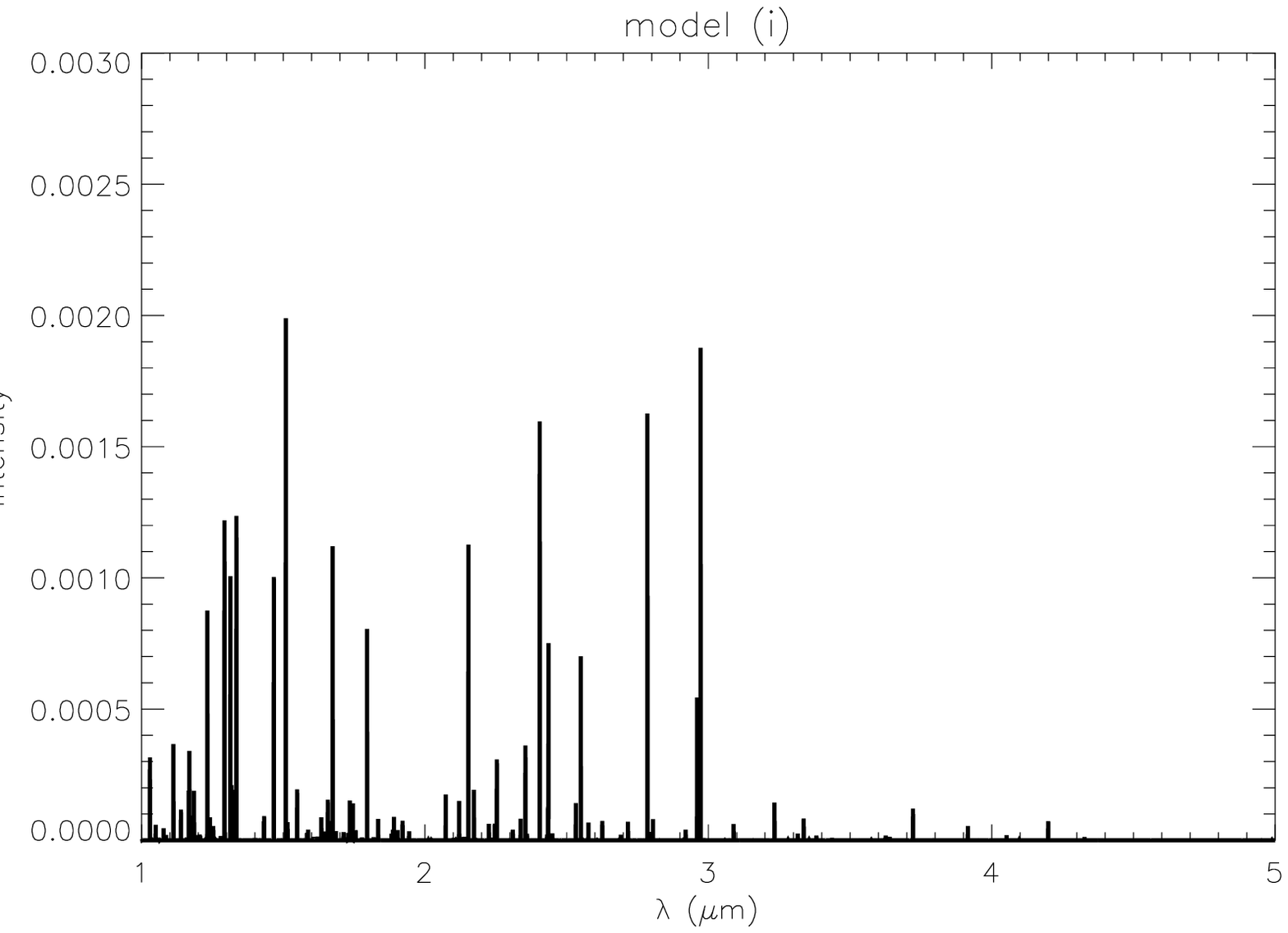}  &
\includegraphics[scale=0.43]{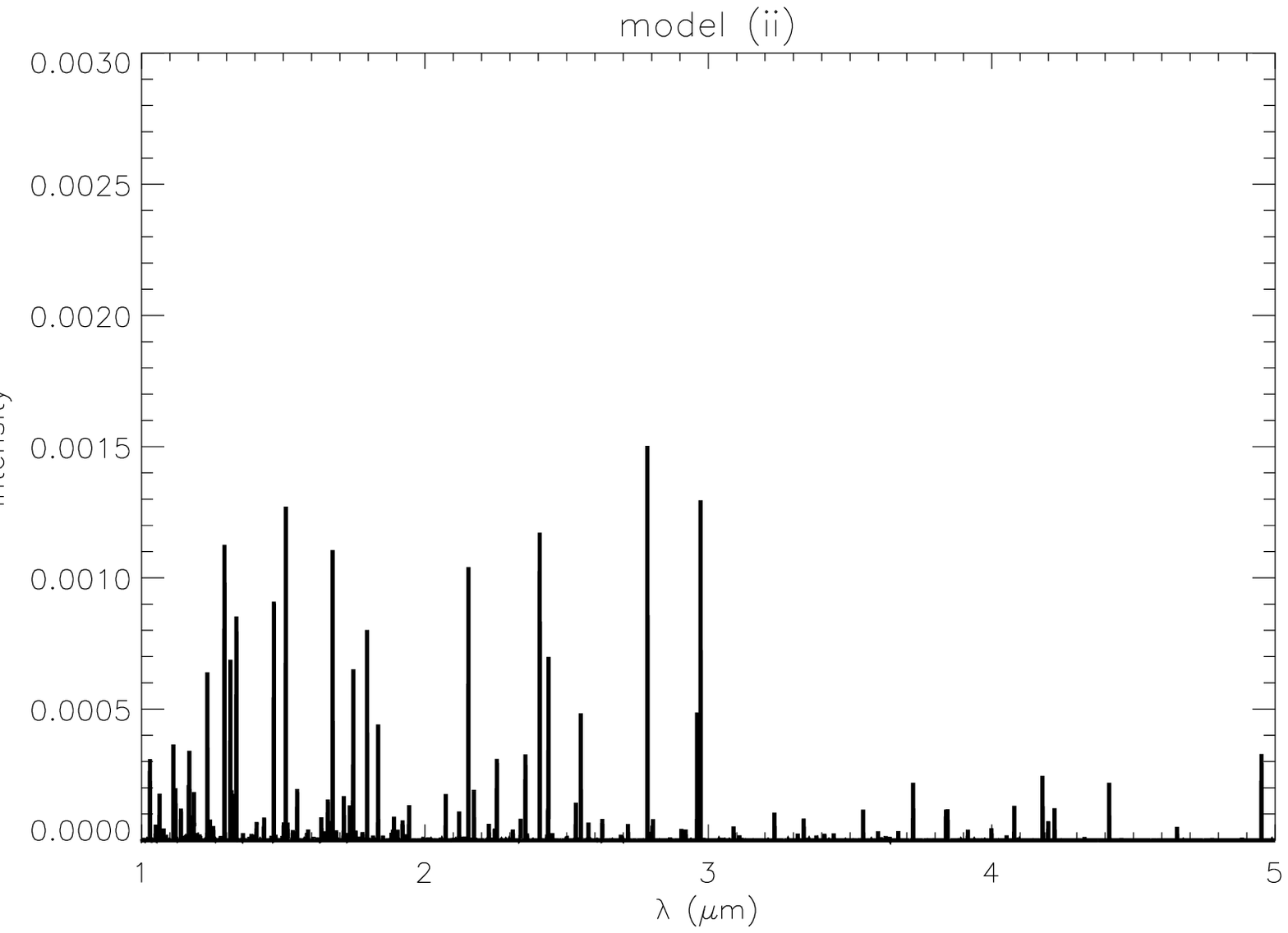} \\
\includegraphics[scale=0.43]{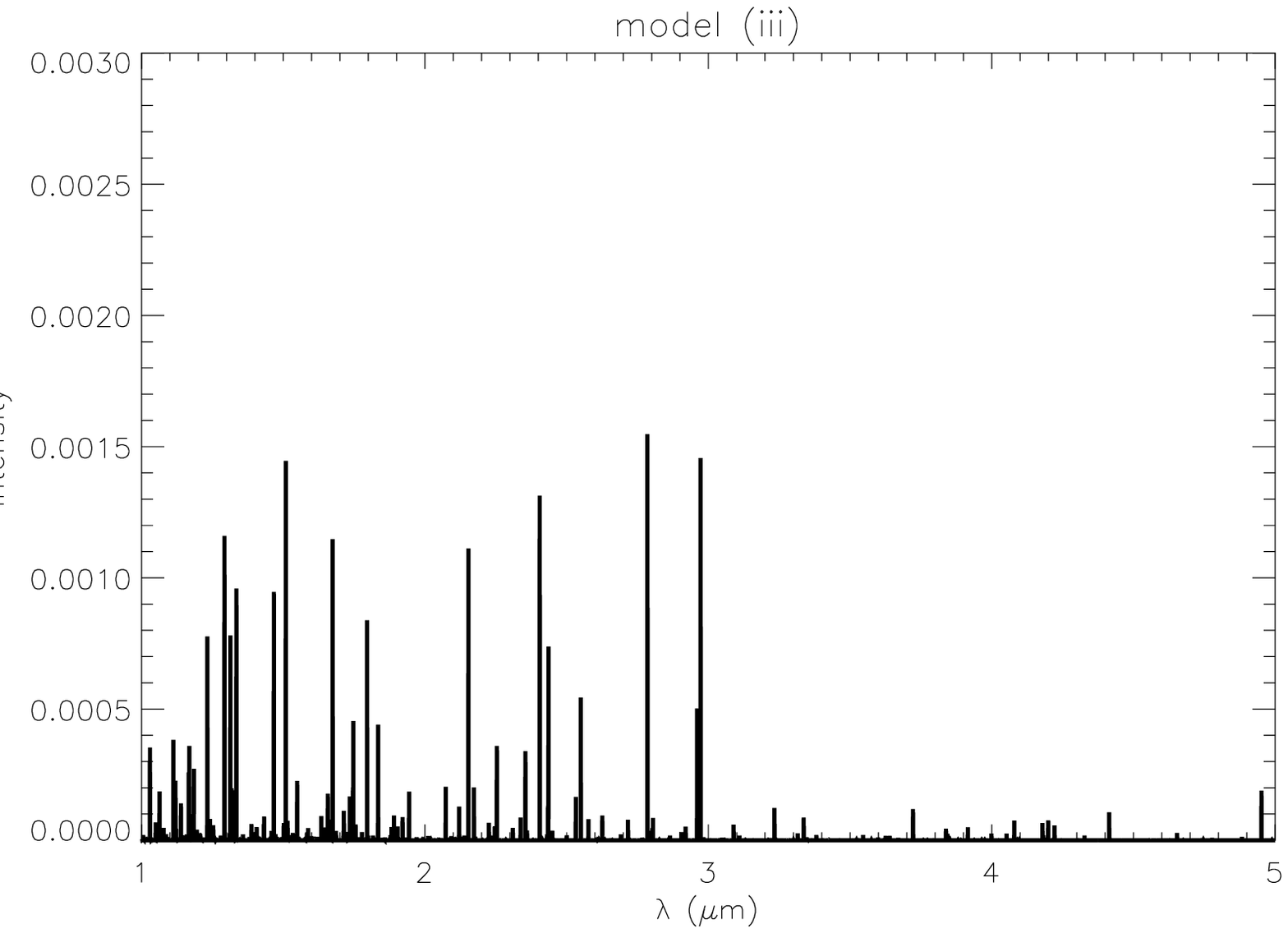}  &
\includegraphics[scale=0.43]{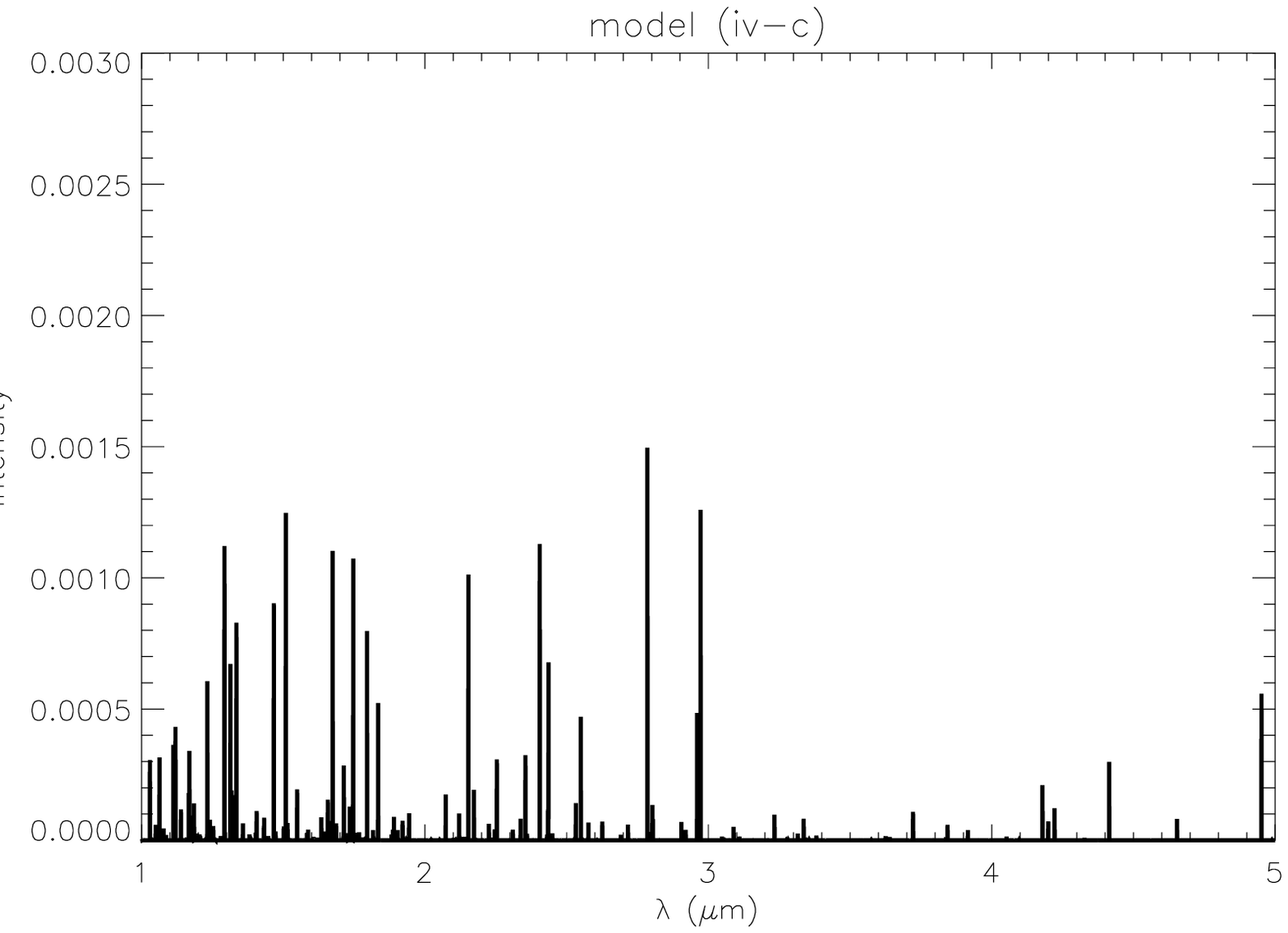} \\
\includegraphics[scale=0.43]{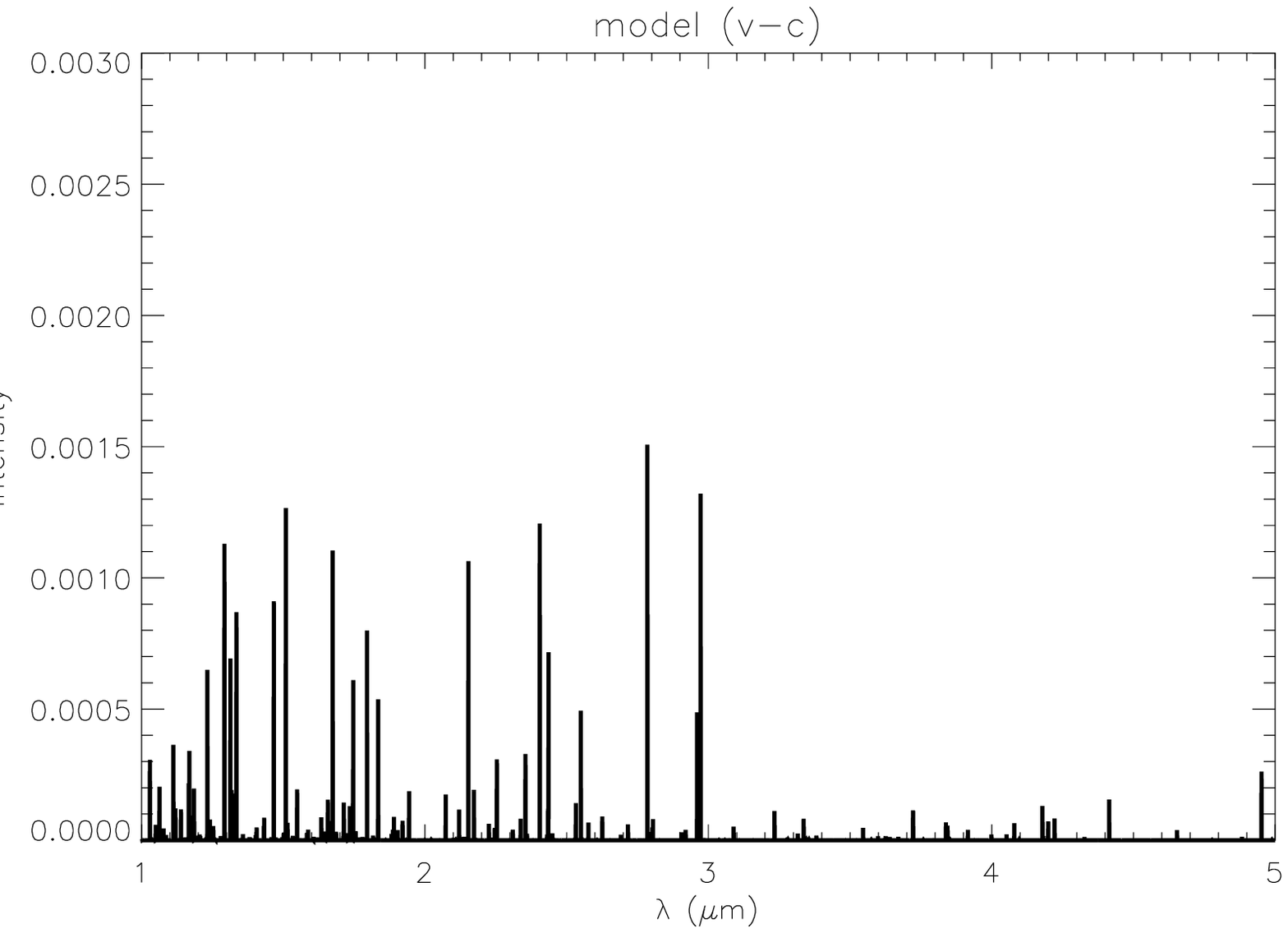} &
\includegraphics[scale=0.43]{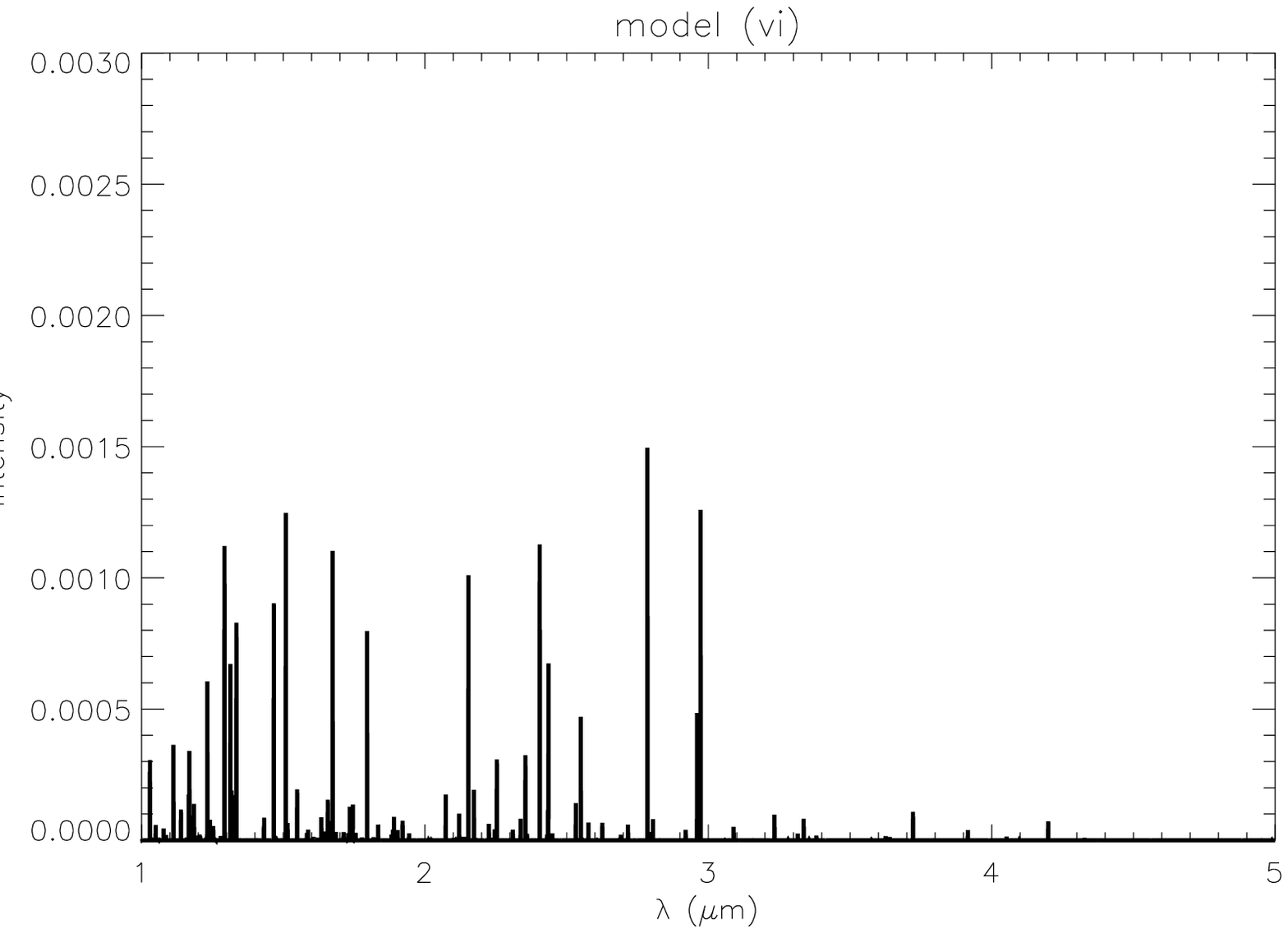}
\end{tabular}
\caption{IR emission in ergs~s$^{-1}$~cm$^{-2}$~sr$^{-1}$~$\mu$m$^{-1}$
computed for the RCM and six different formation pumping models: $(i)$
the model proposed in this work (Table~1); $(ii)$ \citet{BD76}; 
$(iii)$ \citet{DB96}; $(iv-c)$ \citet{TU01} ~\textendash~ carbon model A;
$(v-c)$ \citet{TU01} ~\textendash~ carbon model B; $(vi)$ no formation
pumping. In the case of the \citet{TU01} models, we show only the spectra for
carbon materials, since all the spectra appear to be very similar regardless
of the chemical composition of grain surfaces.}  
\label{fig2}
\end{figure}

\begin{figure}
\includegraphics[scale=0.8]{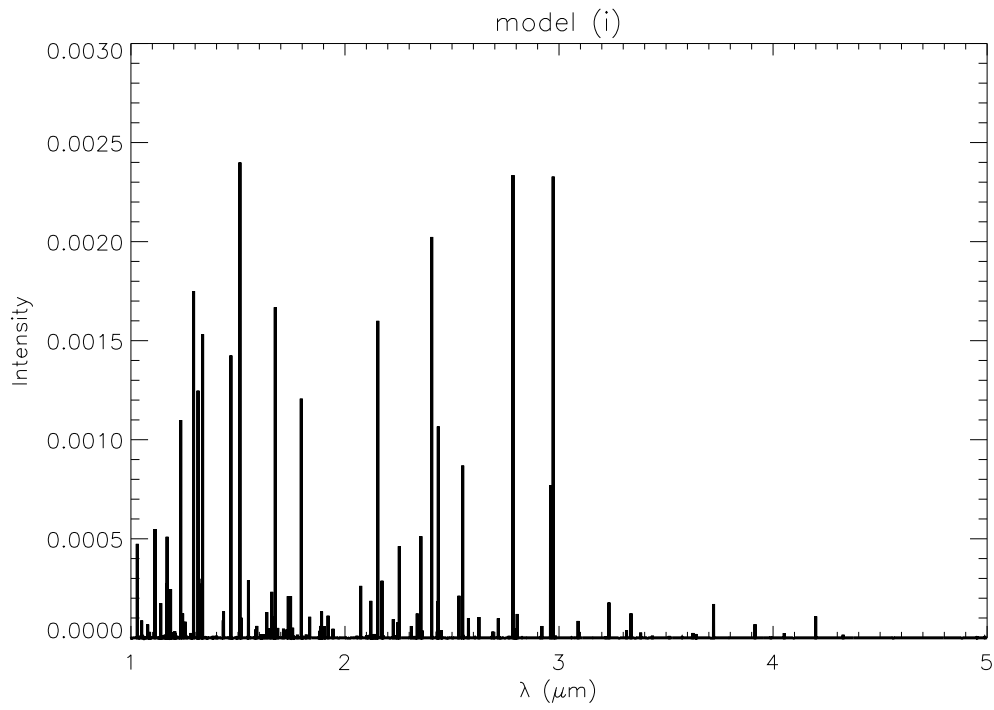} 
\caption{IR emission in ergs~s$^{-1}$~cm$^{-2}$~sr$^{-1}$~$\mu$m$^{-1}$
with formation pumping model $(i)$, computed for the RCM but with visual 
extinction reduced to $A_V = 5$~mag.}\label{fig3}
\end{figure}

In Fig. \ref{fig2}, we compare the results for the six formation pumping
profiles, described in Section~\ref{models}, for the case of dark clouds of 
hydrogen density $n_{\rm H} = 10^6$~cm$^{-3}$, ${\cal R} = 3 \times 
10^{-17}$~cm$^3$~s$^{-1}$, $\zeta = 3 \times 10^{-17}$~s$^{-1}$, standard UV 
field $(\chi = 1)$ and integrated visual extinction $A_V = 100$~mag, hereafter
called the reference cloud model (RCM). The emissivities roughly scale with
the simulated instrumental resolution. We assume a
FWHM of $5 \times 10^{-5} \, \mu$m, approximately corresponding to the 
instrument resolution of echelle spectrometers used by current ground based
telescopes such as the Phoenix instrument on the Gemini telescope, which has a
resolution of $50000 - 80000$ for near\textendash IR wavelengths $1 - 5 \,
\mu$m \citep{H00}. All of these pumping models provide distinct spectral 
features. Therefore, in principle, astronomical observations could be used to 
identify the actual formation pumping mechanism taking place in the ISM. 
However, a common pattern is apparent in all the generated spectra in Fig. 
\ref{fig2}. As the adopted pumping mechanisms result in quite different H$_2$ 
level excitation rates, the underlying common level distribution must be 
generated by some global ambient mechanism, such as thermal pumping. However, 
collisional excitation cannot be significant as the RCM kinetic temperature is 
approximately 10~K in most parts of the cloud. These spectra are the result of 
radiative transfer throughout the whole cloud, enfolding both translucent and 
dense cloud regimes. Therefore, the features present in the IR spectra might 
originate via UV pumping close to the edge of the cloud, in a region in which 
all the hydrogen is not yet in molecular form. In the bottom right panel of 
Fig. \ref{fig2}, we show the IR spectrum generated by model $(vi)$, in which 
formation pumping is not active, but nevertheless a rich spectrum is present: 
radiative pumping in the translucent regime tends to dominate the emission even
at large extinctions. This interpretation is supported by the spectrum shown in
Fig.~\ref{fig3}, derived for a cloud model with the same physical parameters as
the RCM, but with visual extinction $A_V = 5$~mag. The increase of the 
emission in the lower $A_V$ cloud model with respect to the RCM is mainly due
to residual UV radiation coming from the opposite edge of the cloud. In 
Table~2, we report wavelength\textendash integrated emission intensities 
produced in different cloud models, including the spectra displayed in Figs. 
\ref{fig2} and \ref{fig3}. Since the residual field scales with the size of 
the cloud (in plane\textendash parallel clouds of constant density), an 
increase in $A_V$ produces a decrease in the emitted integrated intensity, 
when the main pumping mechanism is radiative. We also note that in the RCM, 
depending on the formation model, H$_2$ formation pumping produces 
an IR excess of about $20 - 40$~\% with respect to the UV\textendash pumped IR
background.

\begin{figure}
\begin{tabular}{cc}
\multicolumn{2}{c}{\includegraphics[scale=0.8]{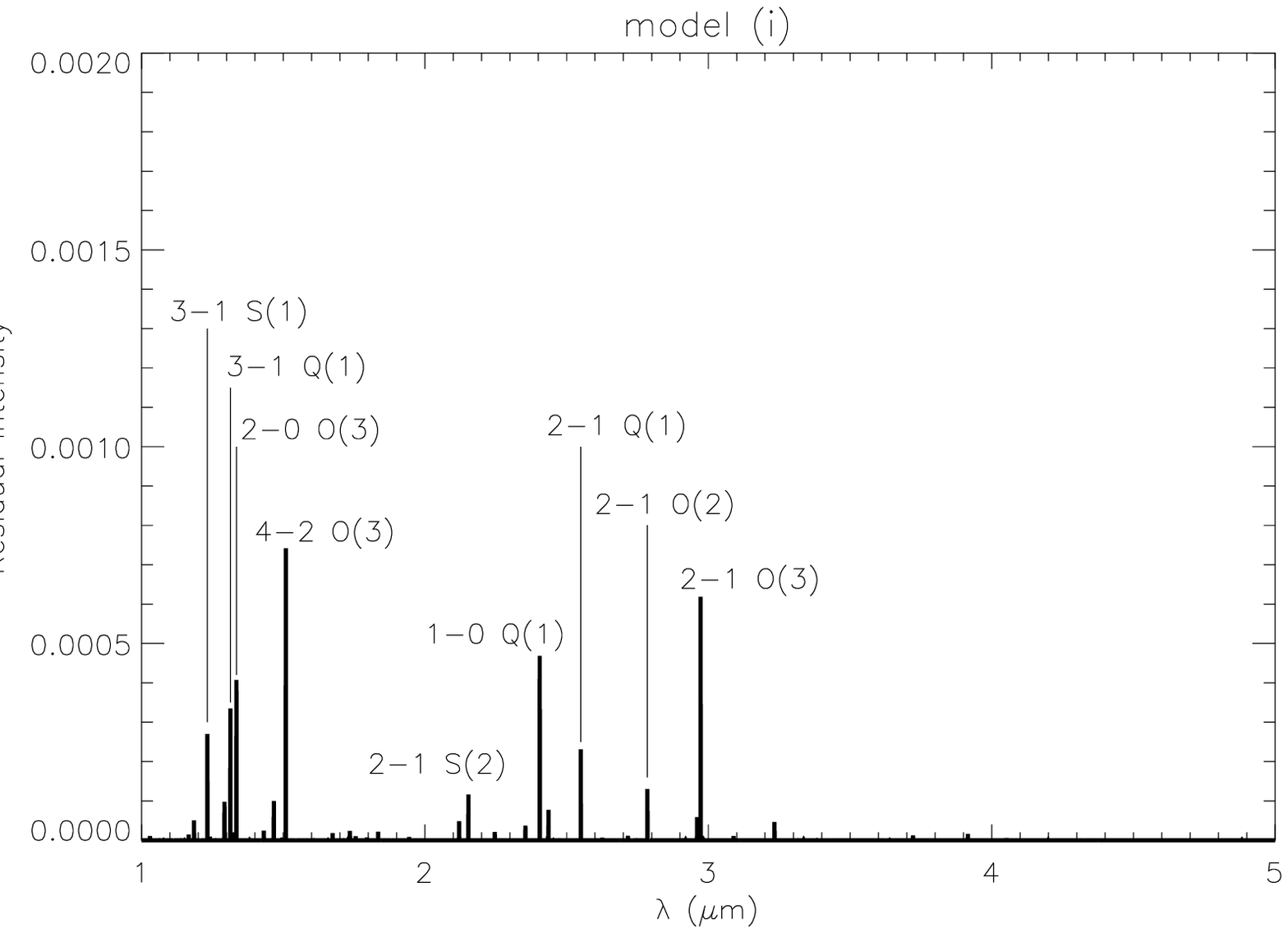}} \\
\includegraphics[scale=0.4]{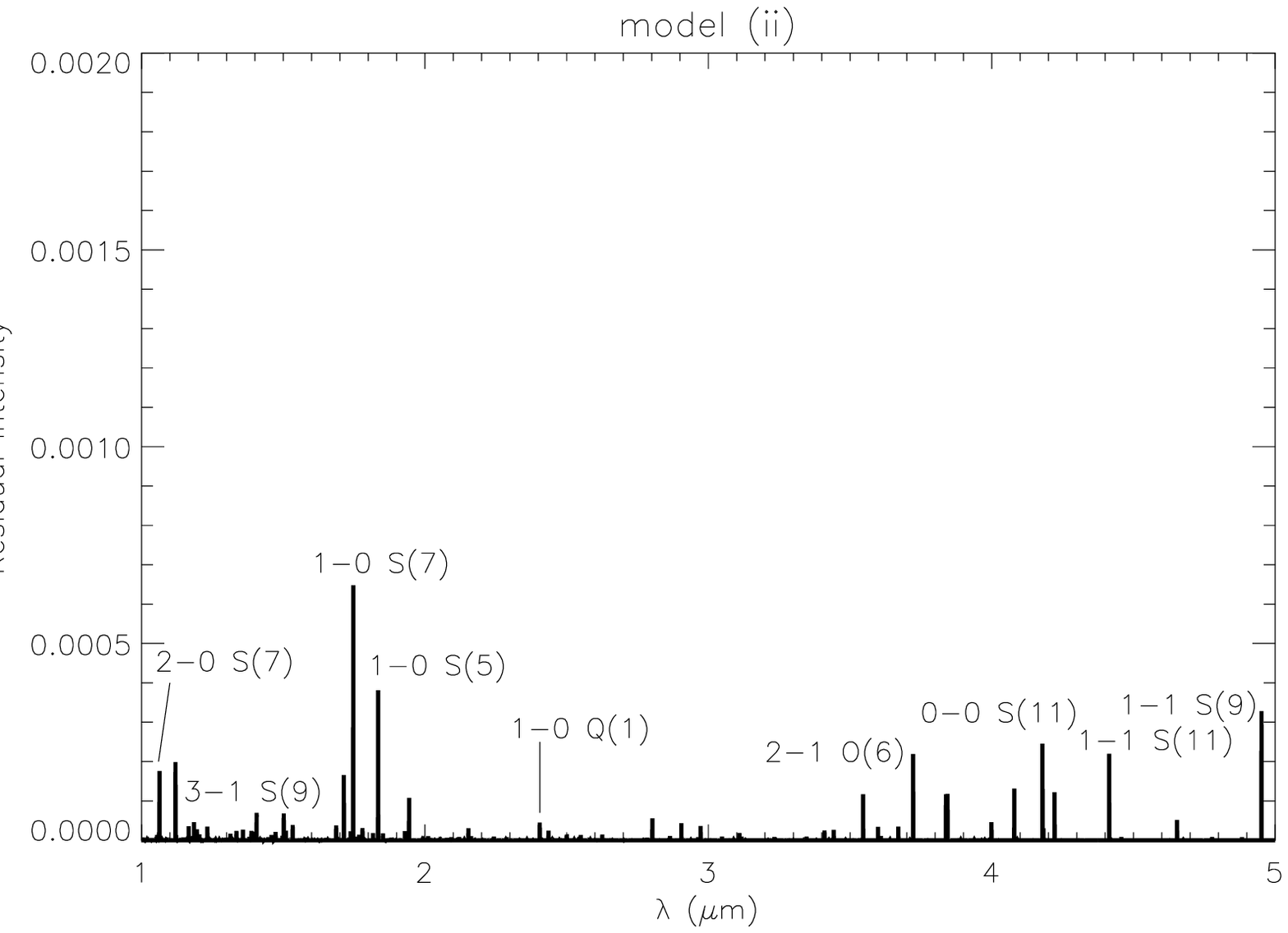} &
\includegraphics[scale=0.4]{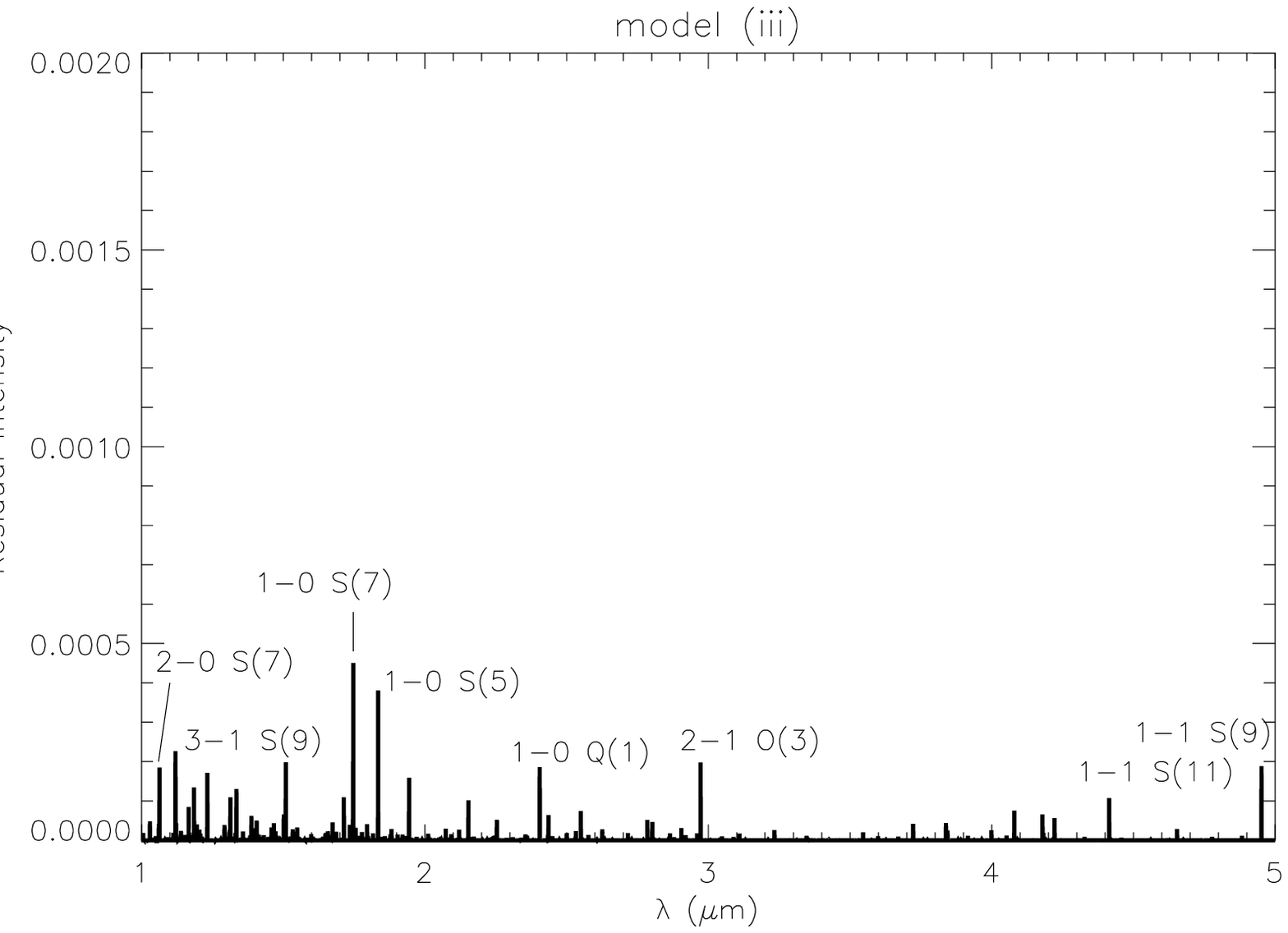}  \\
\includegraphics[scale=0.4]{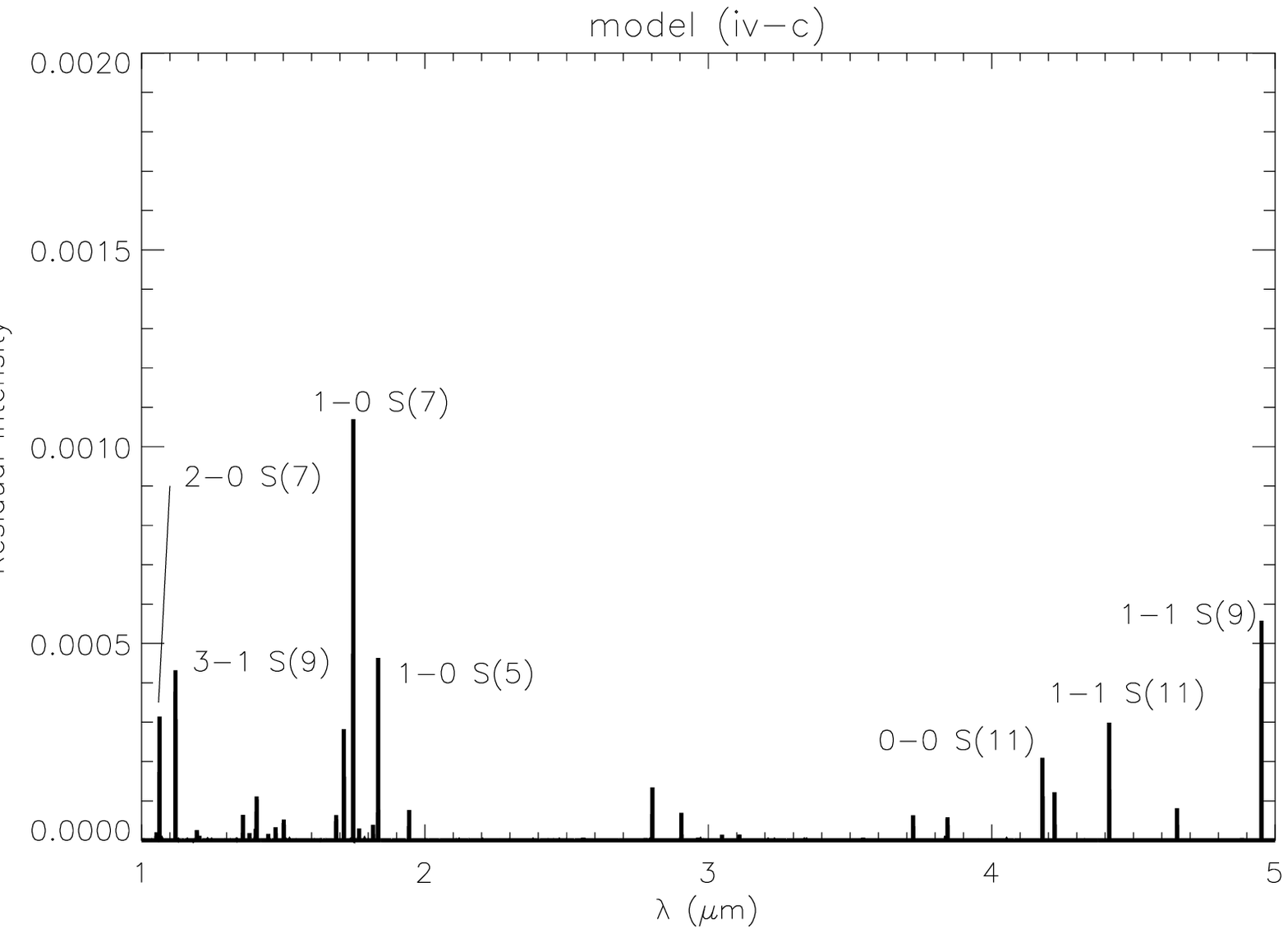} &
\includegraphics[scale=0.4]{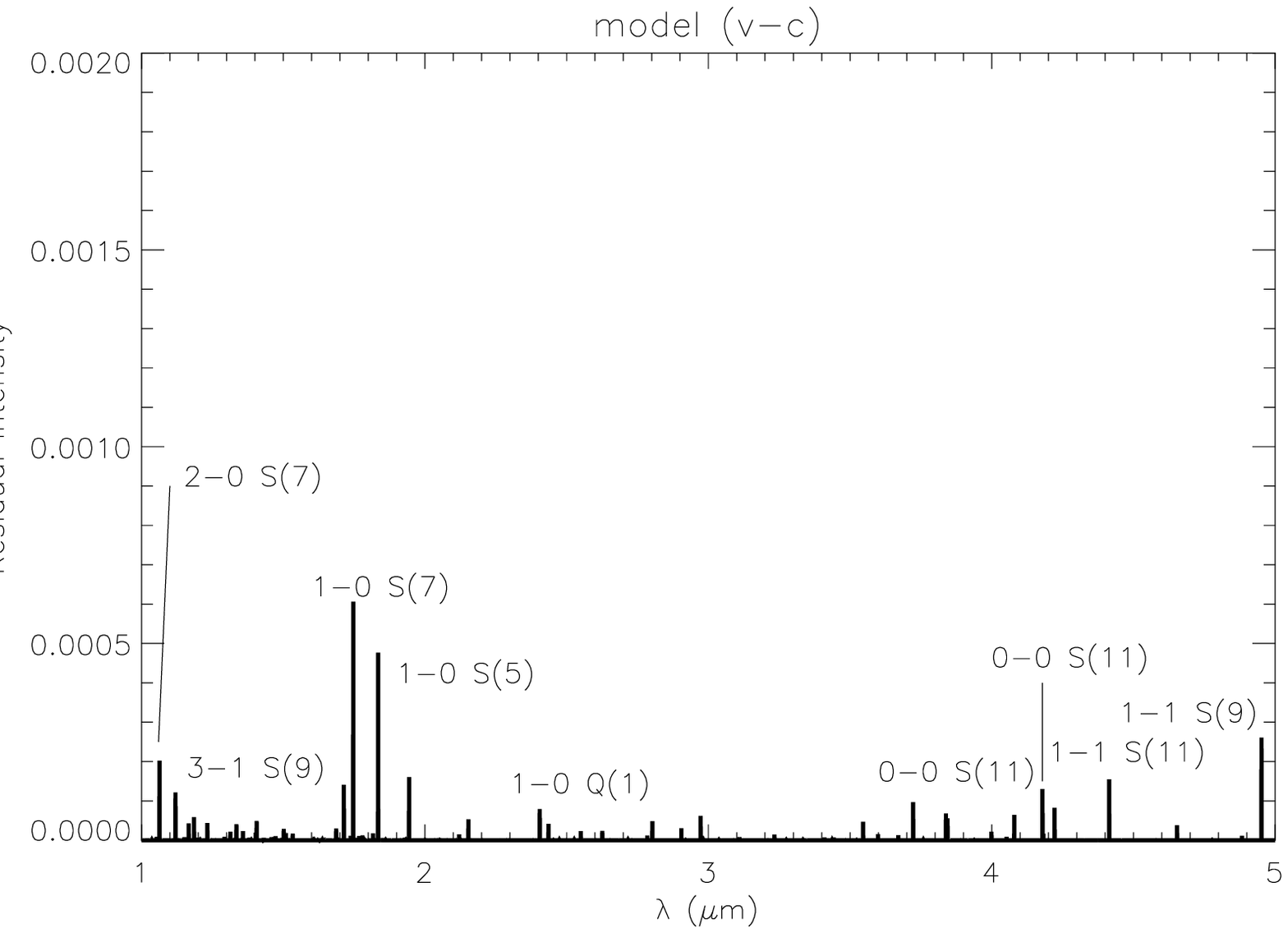} \\
\end{tabular}
\caption{Residual IR spectra. The strongest spectral lines are identified. 
Different models are indicated as in Fig. 2.}
\label{fig4}
\end{figure}

To highlight spectral features arising from specific H$_2$ formation pumping
profiles we show in Fig. \ref{fig4} the residual spectra computed for the RCM
after subtraction of the ``background" UV pumping contribution, i.e. the 
spectrum arising in model $(vi)$  (right bottom panel of Fig. \ref{fig2}). 
Since radiative transfer couples different parts of the cloud, residual 
spectra may provide an indication of the formation pumping effect, but in 
general they cannot be considered as an ``exact" measure of the formation 
pumping contribution to line excitation. The most intense residual emission 
lines produced by the formation pumping are reported for each model in 
Table~3. Model $(i)$ shares some features with model $(iii)$ only. All the 
other formation pumping present several common features. Our proposed 
formation pumping model produces spectra where the highest vibrational level 
is $v = 4$, with no lines from high rotational states in the $4 - 5$ $\mu$m 
region. In model $(ii)$ \citep{BD76}, high rotational levels are pumped, as 
expected, with the highest being $J = 16$ (at an intensity level of $1 \times 
10^{-4}$~ergs~s$^{-1}$~cm$^{-2}$ sr$^{-1}$ $\mu$m$^{-1}$). Although presenting
the highest integrated intensities, models $(ii)$ \citep{BD76} and $(iii)$ 
\citep{DB96} do not show very prominent spectral features, because of the wide
dispersion of the internal energy over a large number of vibrational states. 
Models $(iv)$ and $(v)$ \citep{TU01} show high rotational states, as well as a
combination of moderately high $v$ and $J$, such as the $(v,J) = (3,9)$ state.
These models provide quite different spectral patterns, with the former 
showing brighter lines, while the latter exhibits a richer spectrum. The most 
intense transition for model $(i)$ is $(4-2)$~O(3), whereas the $(1-0)$~S(7) 
line is strongest for all other models. In Table~3, we present emission 
lines computed for model $(vi)$. Since, in this model, formation pumping is 
suppressed, spectral features arise from radiative and collisional pumping 
close to the edge of the cloud. Transitions involving low rotational states 
$(J \le 5)$ show a systematic mixing of internal and environmental pumping
mechanisms. As a consequence, our formation pumping model, which predicts 
little rotational excitation, appears to produce an emission spectrum 
contaminated by external factors. This contamination does not occur for cases 
where the emission is dominated by transitions from high rotational states, 
such as the $(1-0)$~S(7) line in the \citet{BD76} model.

\begin{figure}
\begin{tabular}{c}
\includegraphics[scale=0.8]{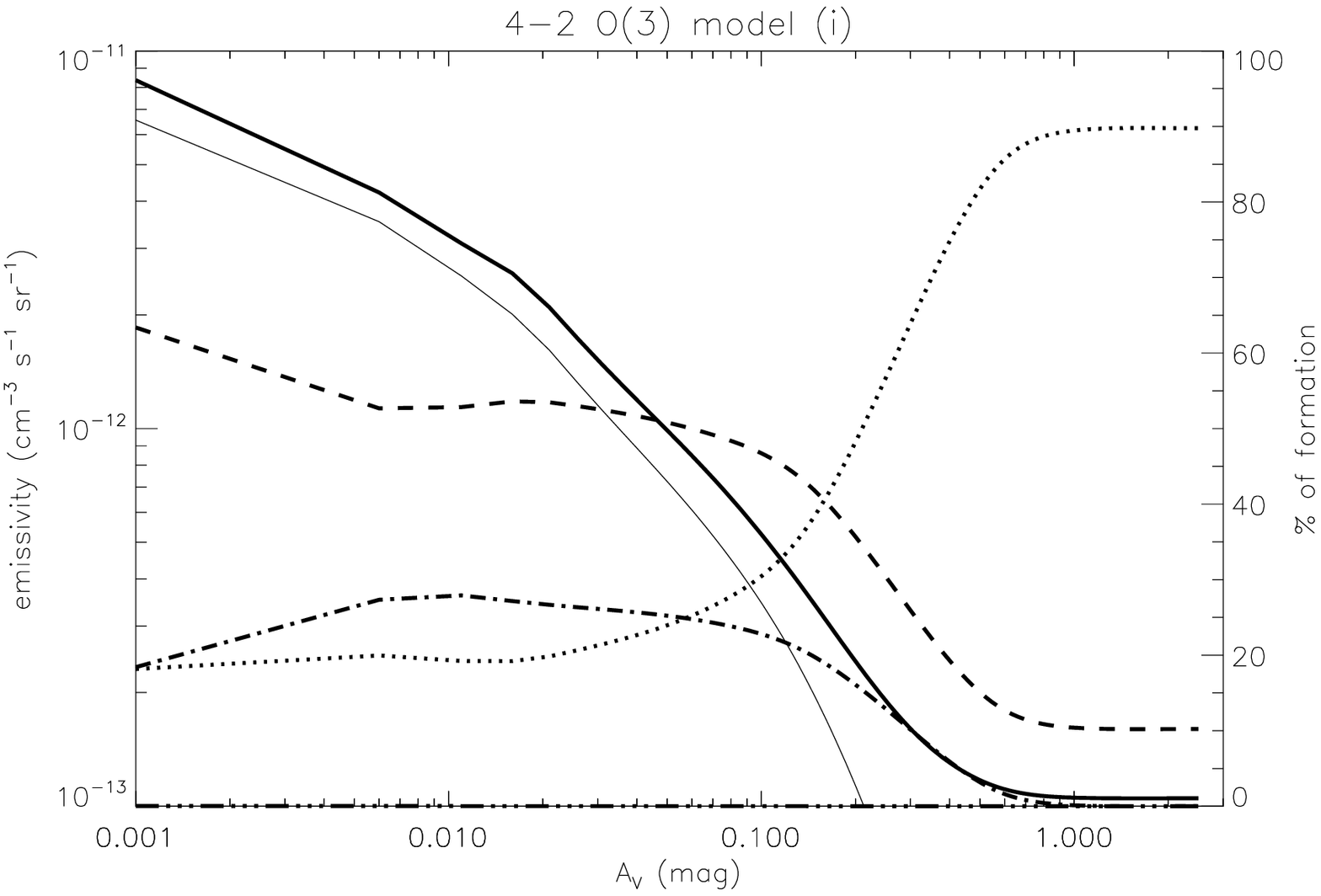} \\
\includegraphics[scale=0.8]{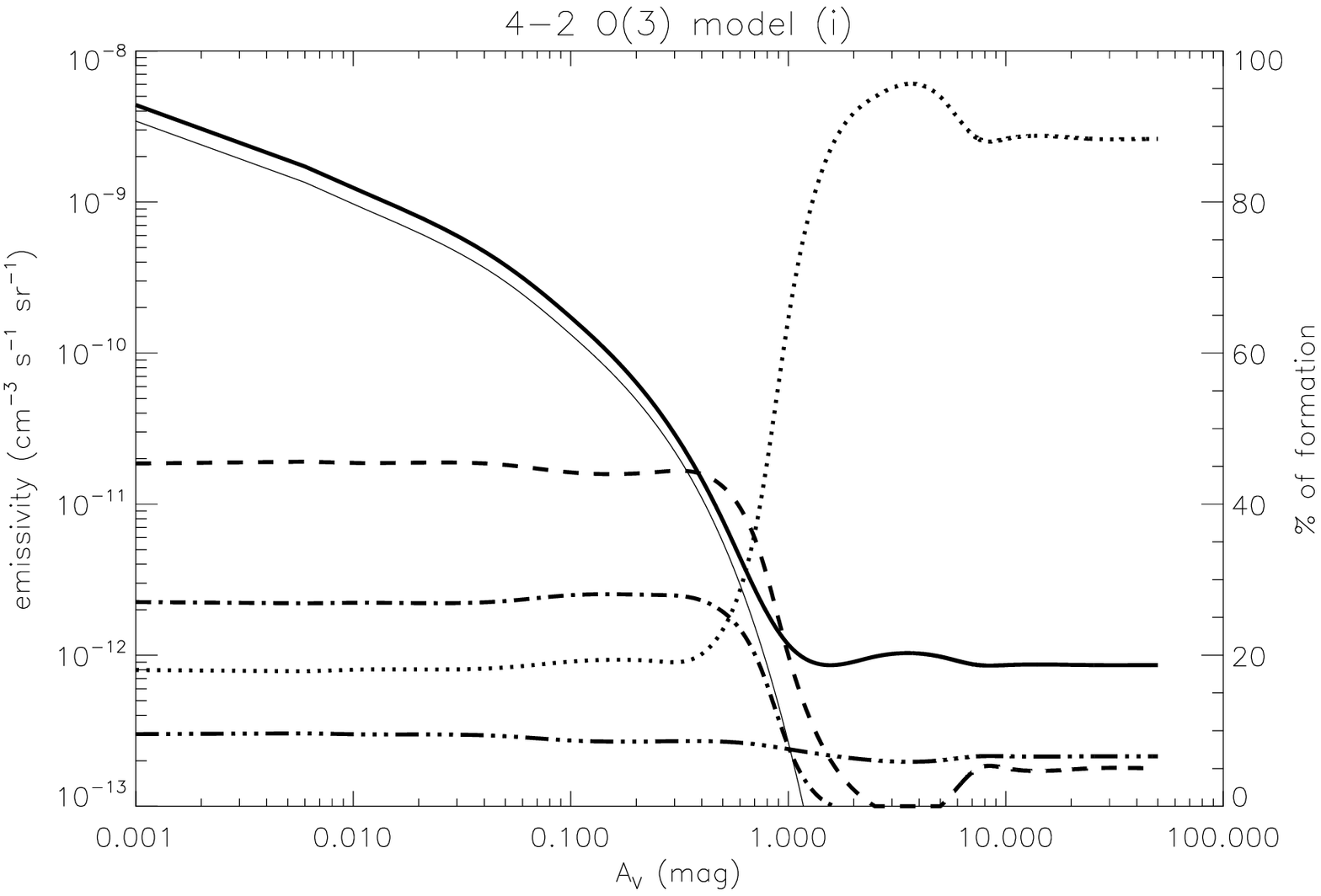}
\end{tabular}
\caption{$4-2$~O(3) line emissivity (cm$^{-3}$~s$^{-1}$~sr$^{-1}$) as a
function of the optical thickness (in mag) computed for model $(i)$ 
(solid thick line) and model $(vi)$ (solid thin line). The relative 
contributions to the population of the upper state of the transition,
$(v,J) = (4,1)$, are: formation pumping (dotted line), IR ro\textendash 
vibrational cascade within the ground electronic state (dashed line),  UV 
fluorescent cascade from excited electronic levels (dot\textendash dashed 
line), and thermal collisions (dot\textendash dot\textendash dot\textendash
dashed line). Upper panel: translucent cloud model 3 (see~Table~2); lower 
panel: RCM.}
\label{fig5}
\end{figure}

\begin{figure}
\begin{tabular}{cc}
\includegraphics[scale=0.43]{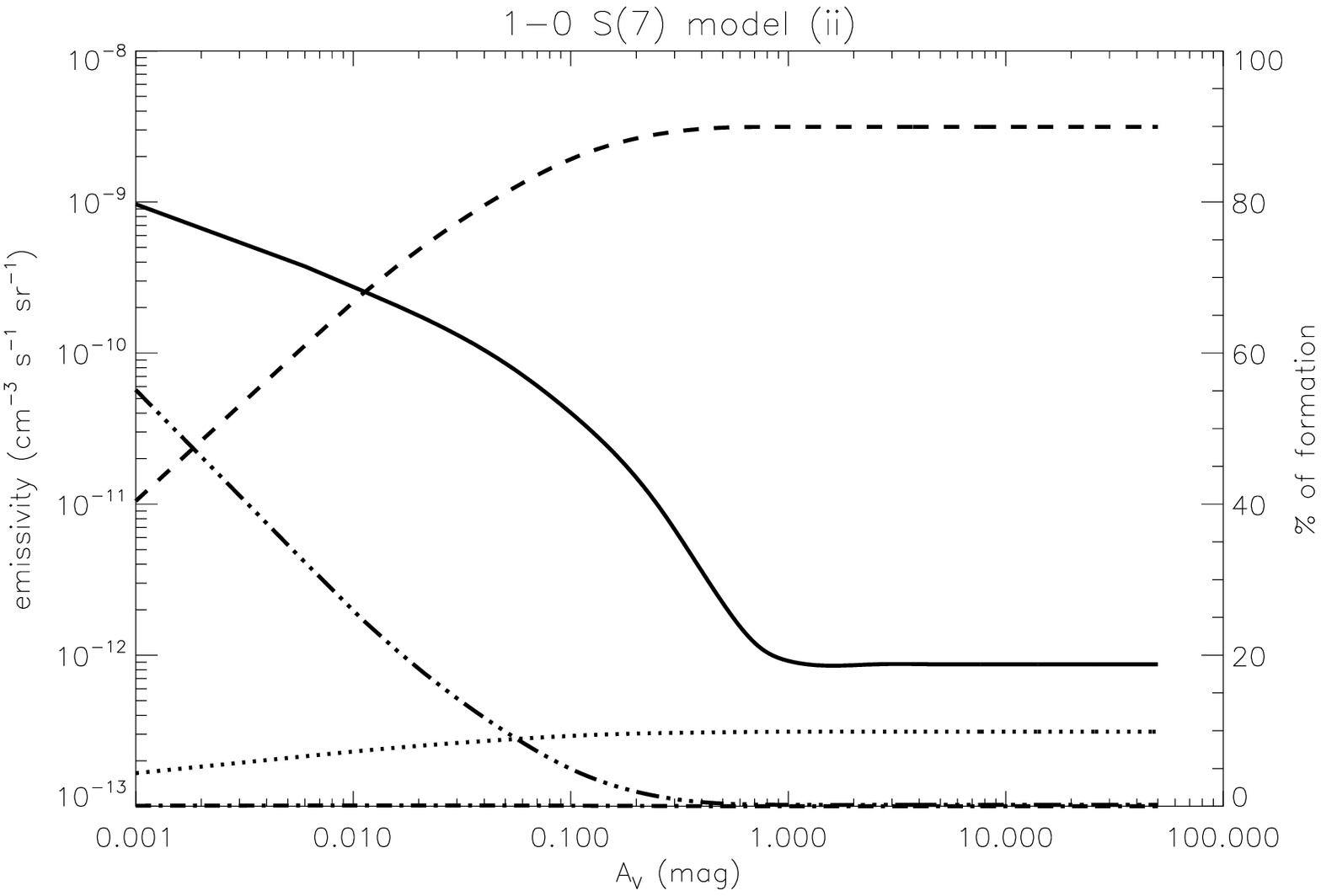} &
\includegraphics[scale=0.43]{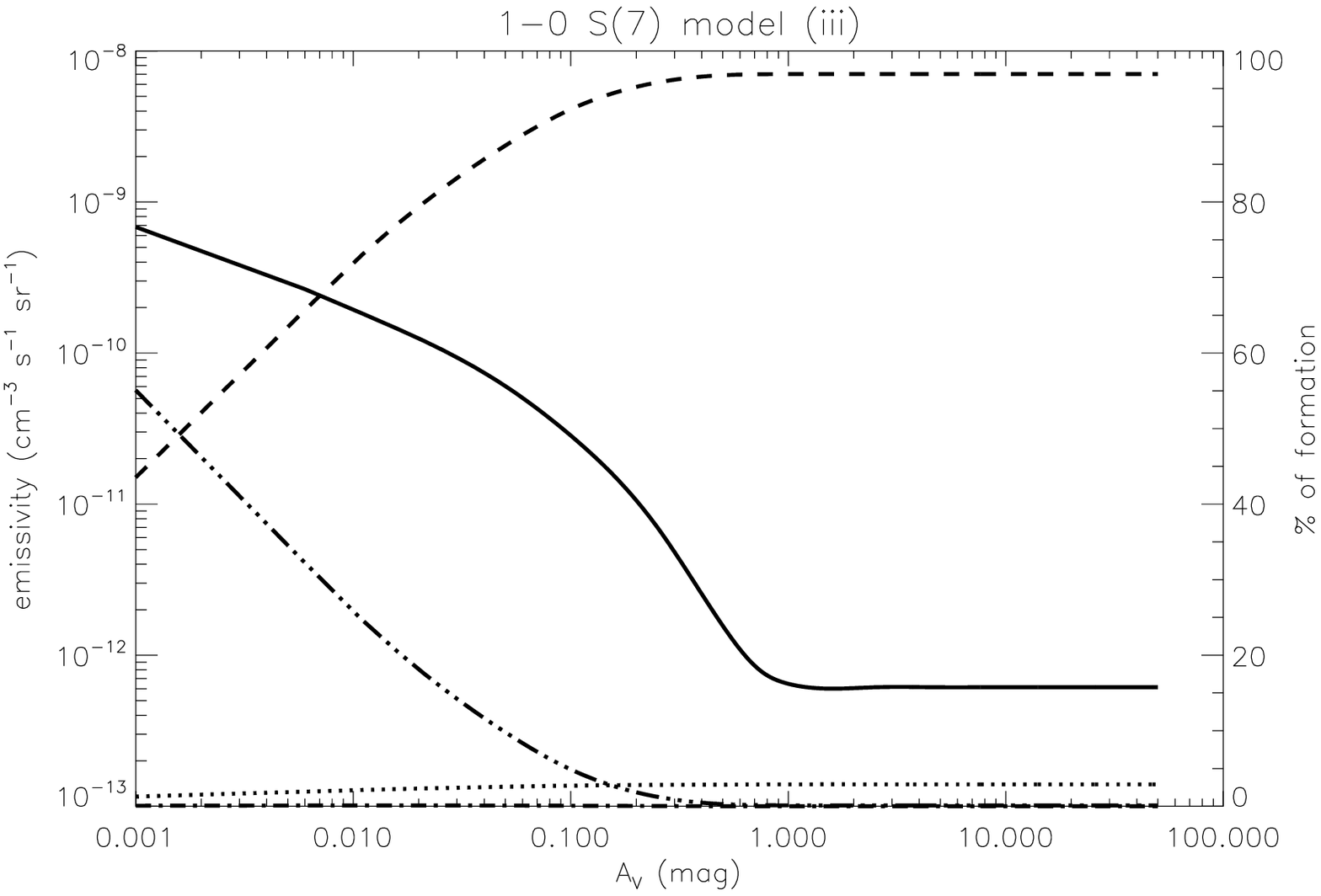}   \\
\includegraphics[scale=0.43]{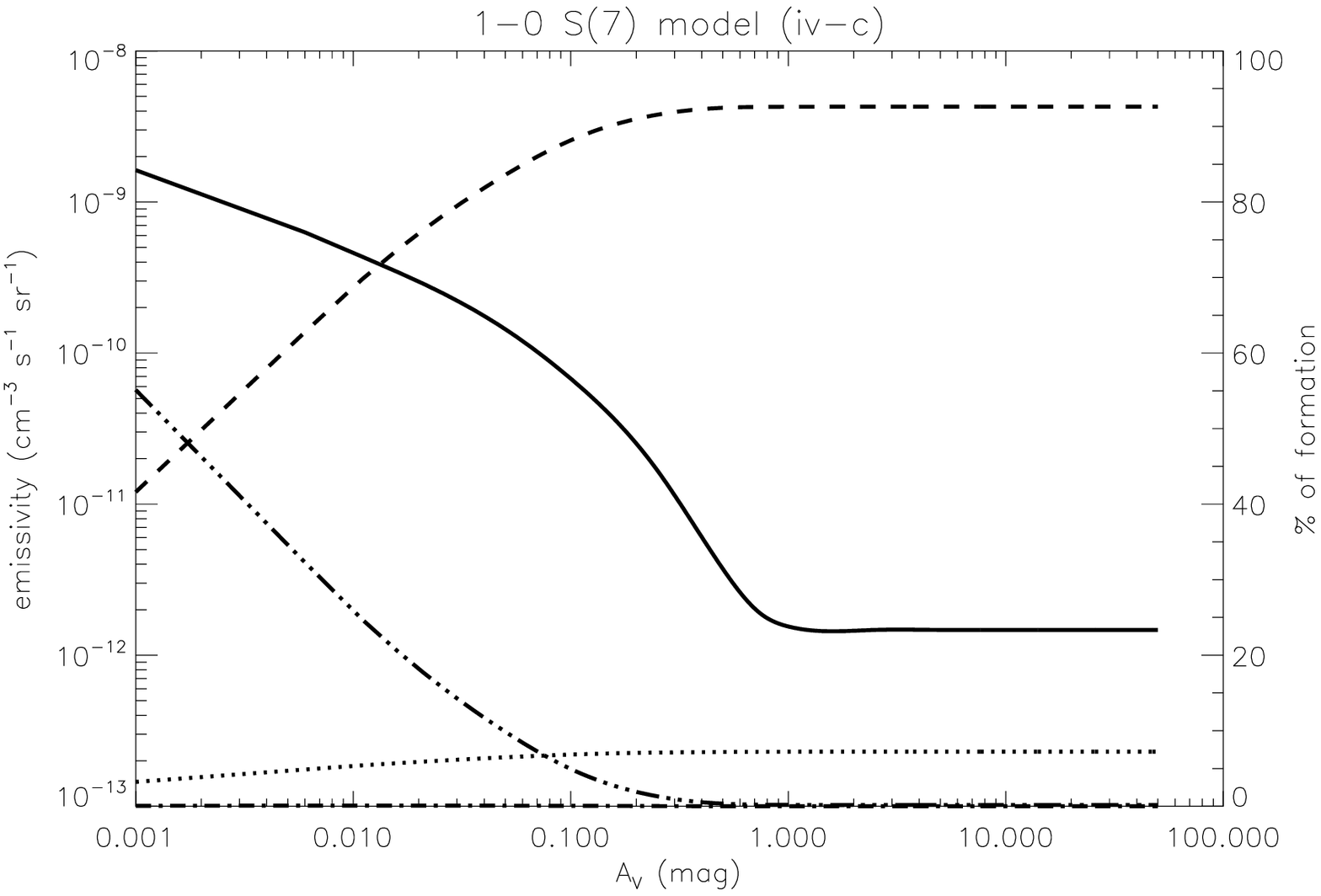} &
\includegraphics[scale=0.43]{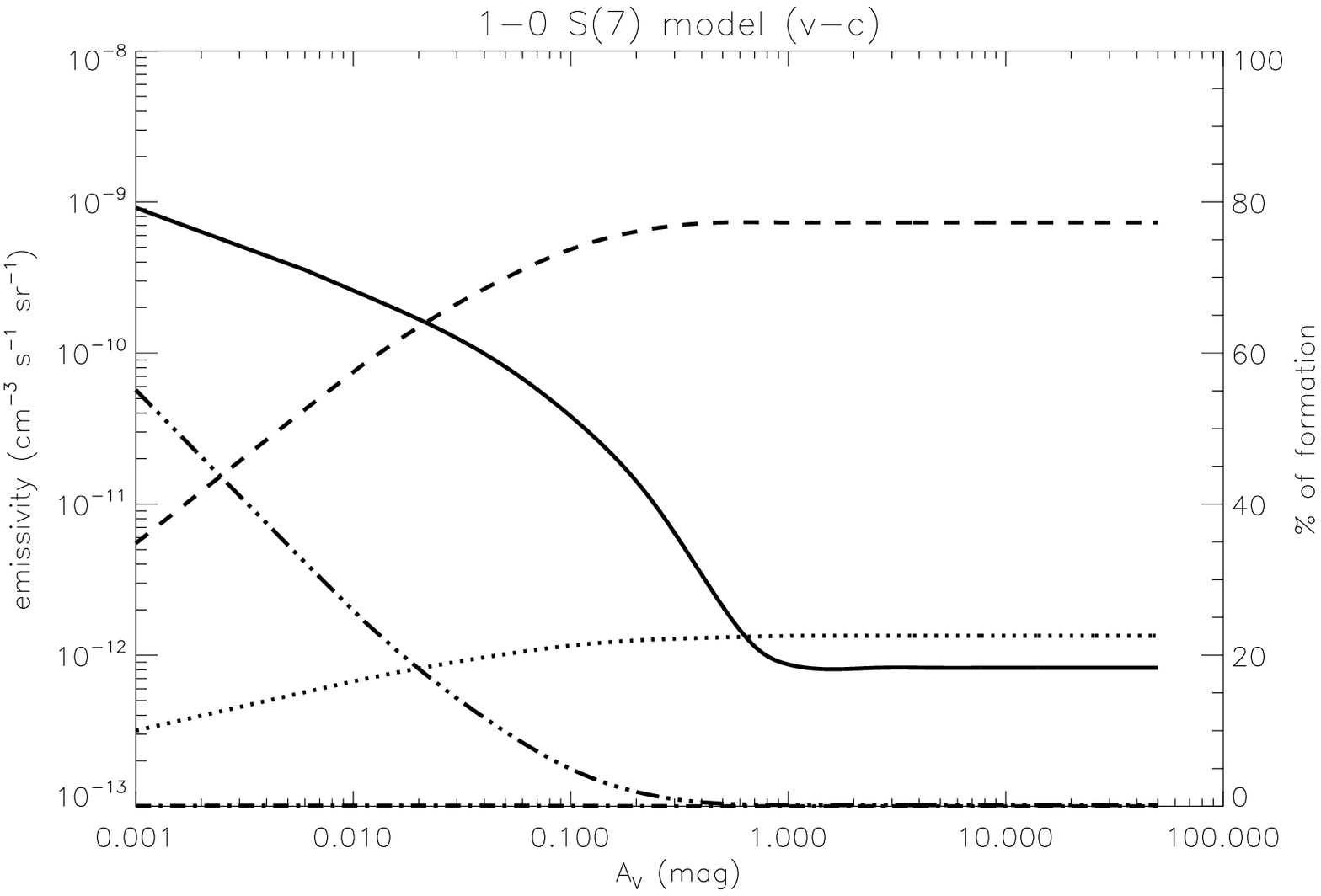} \\
\end{tabular}
\caption{$1-0$~S(7) line emissivities (cm$^{-3}$~s$^{-1}$~sr$^{-1}$) as
functions of the optical thickness (in mag) computed for models $(ii)$,
$(iii)$, $(iv-c)$, and $(v-c)$ (solid lines), in the case of RCM. The relative 
contributions to the population of the upper state of the transition,
$(v,J) = (1,9)$, are: formation pumping (dotted line), IR ro\textendash 
vibrational cascade within the ground electronic state (dashed line),  UV 
fluorescent cascade from excited electronic levels (dot\textendash dashed 
line), and thermal collisions (dot\textendash dot\textendash dot\textendash
dashed line).}
\label{fig6}
\end{figure}

In Figs. \ref{fig5} and \ref{fig6}, we show the volume emissivities of 
the $(4-2)$~O(3) and $(1-0)$~S(7) lines as functions of the optical thickness 
within the cloud. The $(4-2)$~O(3) transition has been computed for models
$(i)$ and $(vi)$ in the translucent and dense cloud regimes, while the 
$(1-0)$~S(7) transition has been computed for models $(ii)$ to $(v)$. In the
figures, we also present the most important relative contributions to the 
population of the upper state of the transitions due to the included 
excitation mechanisms, see  eq.(\ref{SEE}); these mechanisms are UV pumping by
fluorescent cascade from excited electronic states, IR cascade within the 
ground state, thermal collisions and excitation due to H$_2$ formation on dust
grains. The excitation of the $(4-2)$~O(3) line is driven by radiation close 
to the edge of the cloud, but is dominated by formation pumping in the cloud 
interior. Since collisional rates scale with density, thermal collisions 
contribute to the line excitation in response to the very high density adopted
for the RCM. In the case of the low\textendash density translucent cloud 
model, thermal excitation is negligible. In model $(vi)$, line emission is 
essentially the same as for model $(i)$ until the external UV radiation 
density declines sharply for $A_V \ga 1$~mag. No line excitation is produced 
because formation pumping is suppressed in model $(vi)$. For the transition 
$(1-0)$~S(7) in models $(ii)$, $(iii)$, $(iv)$ and $(v)$, the situation is 
different. The line excitation at large visual depth is almost completely 
dominated by the IR cascade ($\sim90~\%$) rather than by direct excitation 
from the formation process ($\sim 10~\%$). However, since UV pumping is not 
effective for large $A_V$, the high energy levels must be populated by 
formation pumping. The IR cascade then populates $(v,J) = (1,9)$, and
lower energy levels. This is 
directly dependent on the distribution of the internal energy over a large 
number of ro\textendash vibrational levels, as occurs in models $(ii)$ to 
$(v)$. In contrast, only a small number of ro\textendash vibrational levels 
are populated in model $(i)$. Therefore, the excitation of the $(1-0)$~S(7) 
line is essentially driven by formation pumping. This conclusion is supported 
by the lack of significant emission at the corresponding wavelength in model 
$(vi)$ (see Table~3).

We now discuss the dependency of the IR spectra on cloud parameters. In 
Table~4, we present line intensities for $4 - 2$~O(3), $1 - 0$~S(7) and 
$1 - 0$~Q(1) transitions using the set of cloud models defined in Table~2.
The physical conditions of such models cover a wide portion of the parameter 
space. In Table~5, we show the excitation and de\textendash excitation rates 
for the upper levels of the $4 - 2$~O(3) and $1 - 0$~S(7) transitions, for 
selected positions within the cloud. We find that 

\begin{itemize}

\item
the integrated spectrum intensity increases almost linearly with $\chi$, as 
expected if the excitation is dominated by the emission in the translucent
regime (see Table~2); the IR excess provided by formation pumping 
decreases with increasing UV radiation density, e.g., in the case of model 
$(i)$ the excess is 20\% when $\chi = 1$, while it is less than 4\% when $\chi
= 1000$;

\item
integrated intensities decrease with increasing $A_V$ (and 
hydrogen column density): this is a direct consequence of the fact that, in a 
double\textendash sided illuminated cloud, the boundaries may be reached by 
residual radiation coming from the opposite edge; integrated intensities do
not decrease when the gas density is $n_{\rm H} = 10^3$~cm$^{-3}$: in such 
a case, since only a partial conversion of hydrogen from atomic to molecular
form takes place, lines pumped by formation get significantly brighter, even
if the overall IR emission does not increase;

\item
line spectra vary marginally with increasing number and column densities 
as soon as hydrogen is almost totally converted to molecular form; in clouds 
where partial H/H$_2$ conversion occurs, line intensities are larger than in 
dense clouds where emission paths are considerably longer; 

\item
the strength of a line increases steadily, although not linearly, with kinetic
temperature, because as the temperature rises, more collisional pumping 
occurs, although the shape of the spectrum does not greatly alter; in general, 
when kinetic temperature increases over 100~K in a substantial fraction of
the cloud, there is a sharp rise in the strength of the $1 - 0$~Q(1) line, due
to thermal excitation of H$_2$ to the $(v,J) = (1,1)$ level; 

\item
for both levels $(v,J) = (1,9)$ and $(4,1)$ in any position within the RCM, 
the major exit channel is radiative de\textendash excitation; the major
entry channel is direct formation pumping for level $(v,J) = (4,1)$ and
IR cascade from upper levels for $(v,J) = (1,9)$; since both UV fluorescence
and thermal collisions are neglible in most parts of the cloud, the pumping of 
high ro\textendash vibrational states is due H$_2$ formation on dust grains;

\item
as there is uncertainty in the population of the $v = 0$ levels (see Section 
2.1), H$_2$ spectra have been generated for a distribution where the $v = 0$ 
levels are twice as populated as the $v = 1$ levels and for a distribution 
where the $v = 0$ levels are half as populated as the $v = 1$ levels; these
changes make only a minimal difference to line strengths, with spectra 
similar to within 4\%.

\end{itemize}

\begin{figure}
\epsscale{.80}
\plotone{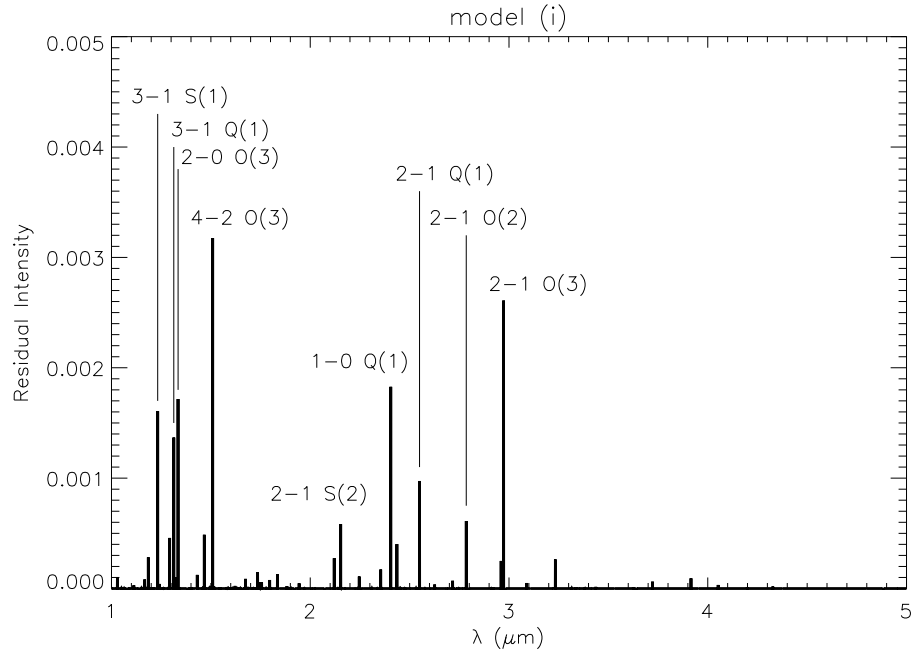}
\caption{Residual IR spectrum for a translucent cloud with gas density 
$n_{\rm H} = 10^3$~cm$^{-3}$ and $A_V = 5$~mag. All other parameters are
as for RCM.}\label{fig7}
\end{figure}

Finally, in Fig. \ref{fig7}, we show the residual emission spectrum of a 
standard translucent cloud ($n_{\rm H} = 10^3$~cm$^{-3}$, $A_V = 5$~mag, 
all other parameters are as for the RCM) for the case of model $(i)$. It 
appears that in a cloud of moderate density and extinction, the intensity of 
the emitted spectrum is $3 - 4$ times more intense than in the case of the RCM,
in which the hydrogen column density is 20 times larger. The increase in line 
strength is more evident for lines produced during H$_2$ formation. This
reflects the much larger abundance of atomic hydrogen (by about a factor of 
1000) in the lower density cloud. The comparison with the RCM case also shows 
that the very long emission path inside a dense cloud produces only a marginal
increase in the line intensities, since the abundances of the upper levels in 
the emitting transitions fall abruptly as soon as H$_2$ formation saturates. 

\section{Discussion and Conclusions}
In this work we investigate the effects of formation pumping in IR H$_2$ 
emission spectra with a new formation pumping model (Table~1). We construct 
radiative tranfer and chemical models for H$_2$ newly\textendash formed on 
dust grains. By using realistic space\textendash dependent cloud models, we 
find that UV radiative pumping dominates the emission even in clouds with very 
high visual extinction. After eliminating radiative pumping, we obtain 
residual IR spectra due to formation pumping. When cloud kinetic temperatures 
rise over 100~K, thermally excited ro\textendash vibrational levels of the 
$v = 1$ manifold may contribute to the overall IR spectrum. 

Surprisingly, spectra show a very modest increase with both volume and column 
densities. IR emission is expected to scale with the square of volume density
and linearly with column density, via the formation rate and the path along 
the line of sight. However, deep within a cloud ($A_V \ga 2.5$~mag) an almost 
total conversion of hydrogen from atomic to molecular form occurs. Thus, the 
emissivity produced in the inner zone of a cloud is ``lost" within the 
contribution originating in the transition zone, in which H$_2$ abundances are 
still comparable to those of atomic hydrogen. In other words, the bulk of the
emission is coming from the ``translucent" regime close to the edge of a cloud.

This can be easily understood considering a simple two\textendash level model
representing an emission line $u \to l$ arising during the H$_2$ formation 
process within a homogeneous cloud. In such a representation, the line 
emissivity is given by 
\begin{equation}\label{emis}
\epsilon = \frac{hc}{4 \pi \lambda} A_{ul} n_u \phi_\lambda
\quad {\rm with} \, A_{ul} n_u \sim {\cal R} n_{\rm H} n_1 \delta_u,
\end{equation}
(see Table~5), while the brightness is obtained by integration of 
eq.(\ref{emis}) along the line of sight $\Delta L$,
\begin{equation}\label{bright}
\delta I_\lambda \sim  \frac{hc}{4 \pi \lambda} {\cal R} n_{\rm H}^2 x_1 
\delta_u \phi_\lambda \Delta L.
\end{equation}
In eqs.(\ref{emis}) and (\ref{bright}), $ \phi_\lambda$ is the assumed line 
profile, $n_u$ is the population of the upper level of the transition and 
$x_1$ is the fractional abundance of atomic hydrogen. The intensity ratio 
between emissions originating from dense and translucent 
regions in the cloud then results
\begin{equation}
\frac {\delta I_\lambda^D}{\delta I_\lambda^T} = \frac {x_1^D N_{\rm H}^D}
{x_1^T N_{\rm H}^T} \la 10^{-3} \times \frac {N_{\rm H}^D}{N_{\rm H}^T}. 
\end{equation}
We get $\delta I_\lambda^D \sim \delta I_\lambda^T$ when $N_{\rm H}^D \sim 1000
\times N_{\rm H}^T$. Since in the RCM, $N_{\rm H}^D / N_{\rm H}^T \sim 20$, we 
finally obtain $\delta I^D \la 2\% \, \delta I^T$. Thus, emission from the 
translucent outer regions of a cloud dominates the spectral line intensity, 
with only a small contribution from the dense central regions of the dark 
cloud. Consequently, models of formation pumping that are rotationally cool, 
such as the one put forward in this work, are heavily contaminated by pumping 
processes arising from nebular physical conditions.

It is possible that only lines from high rotational levels can be 
identified observationally as being due to formation pumping. Unfortunately,
most of the models proposed in literature have common features (see Table~3) 
and, in general, it appears difficult to discriminate between them. In 
contrast, our new formation pumping model produces a spectrum differing 
significantly from the line patterns of all the other proposed models, and may
provide an unambiguous signature for detection, but only from regions not 
contaminated by UV radiation. In addition, the observation of H$_2$ molecules 
newly\textendash formed on dust grains appears to be currently  very 
difficult. The peak intensity of the spectral lines are of the order of 
$0.001$~erg~s$^{-1}$~cm$^{-2}$~sr$^{-1}$~$\mu$m$^{-1}$ (see Fig. 2), or 
$I_\lambda \sim 1 \times 10^{-16}$~W~m$^{-2}$~arcsec$^{-2}$~$\mu$m$^{-1}$. 
Using a resolution $R =  \lambda /\Delta \lambda = 37000$ (taken from UKIRT 
webpages\footnote{ 
http://www.jach.hawaii.edu/UKIRT/instruments/cgs4/optical/resolution.html})
we obtain an integrated brightness at $2 \, \mu$m of $I \sim 1.2 \times 
10^{-21}$~W~m$^{-2}$~arcsec$^{-2}$. Hence, using the UKIRT 1\textendash pixel 
slit, which is 0.609~arcsec wide, we obtain a flux of $F \sim 4.5 \times 
10^{-22}$~W~m$^{-2}$ (for an extended source). This flux is much lower in 
comparison to the $3\sigma$ 30\textendash minute sensitivity per pixel 
obtained with the echelle grating, which has values $8 \times 
10^{-20}$~W~m$^{-2}$ at $1.6 \, \mu$m and $6 \times 10^{-20}$~W~m$^{-2}$ at 
$2.2 \, \mu$m. As a consequence, dark clouds are not a good place to look for 
signatures of H$_2$ formation, since an increase in the emission path does not
correspond to a significant increase in the intensity of the spectrum. 

Therefore, H$_2$ formation pumping may be undetectable in those regions in 
which molecular hydrogen is not destroyed at a fairly fast rate. Thus, within 
the current regime of instrument sensitivity, the non\textendash detections of
H$_2$ IR emission in dense clouds reported by \citet{T03} and \citet{C09} do 
not need to be explained by the thermalization of nascent H$_2$ molecules on 
the surface of dust grains, as suggested by \citet{C09}. In dense PDRs, UV 
radiation maintains a substantial level of atomic hydrogen in the gas. 
However, the radiation also causes excitation of H$_2$ to the Lyman and 
Werner bands, inducing IR fluorescence to further complicate the emission 
process. Much more suitable regions for observing formation pumping appear to 
be X-ray dominated regions (XDRs). High energy X\textendash rays penetrate 
much deeper into gas clouds than UV photons.  X\textendash rays are
preferentially absorbed by heavy elements to produce multiply\textendash 
charged ions and photo\textendash ionize the gas deeply within a cloud. High 
energy primary photo\textendash electrons deposit their energy into the gas, 
inducing a secondary electron cascade, which ionizes, excites and dissociates 
atomic and molecular species, and also heats the gas through Coulomb 
collisions. Moreover, although the electrons efficiently destroy molecular 
hydrogen, electron discrete interactions only provide excitations in the H$_2$ 
vibrational ladder up to $v = 2$ \citep{D99}. The radiative decays of the 
electronically excited H and H$_2$ produces FUV photons, H$_2$ 
Lyman\textendash Werner photons and H Ly$\alpha$ photons. We can estimate the 
impact of photo\textendash electron induced UV radiation in the following
way: assuming photo\textendash electrons of mean energy 30~eV (as in 
\citealt{T97}), the number of excitations to the states $B^1\Sigma_u^+$ and 
$C^1\Pi^u$, including the contribution of cascading from higher singlet 
states, are approximately 0.6 and 0.4, respectively \citep{D99}. The 
excitation rate to the excited electronic states is thus $\sim \zeta_X$, where
$\zeta_X$ is the total X\textendash ray ionization rate. The radiative 
excitation of Lyman and Werner bands is roughly $5 \times 10^{-10} \, 
\chi$~s$^{-1}$ \citep{DB96}. We obtain the equivalent UV field scaling factor 
by means of the relation $\chi = \zeta_X / 5 \times 10^{-10}$. Thus, the most 
prominent feature in the IR spectrum generated using our formation pumping 
model, namely the transition $4-2$~O(3), would be minimally affected by 
X\textendash rays, as long as the ionization rate is $\zeta_X \ll 
5 \times 10^{-10}$~s$^{-1}$.  

\begin{figure}
\epsscale{.80}
\plotone{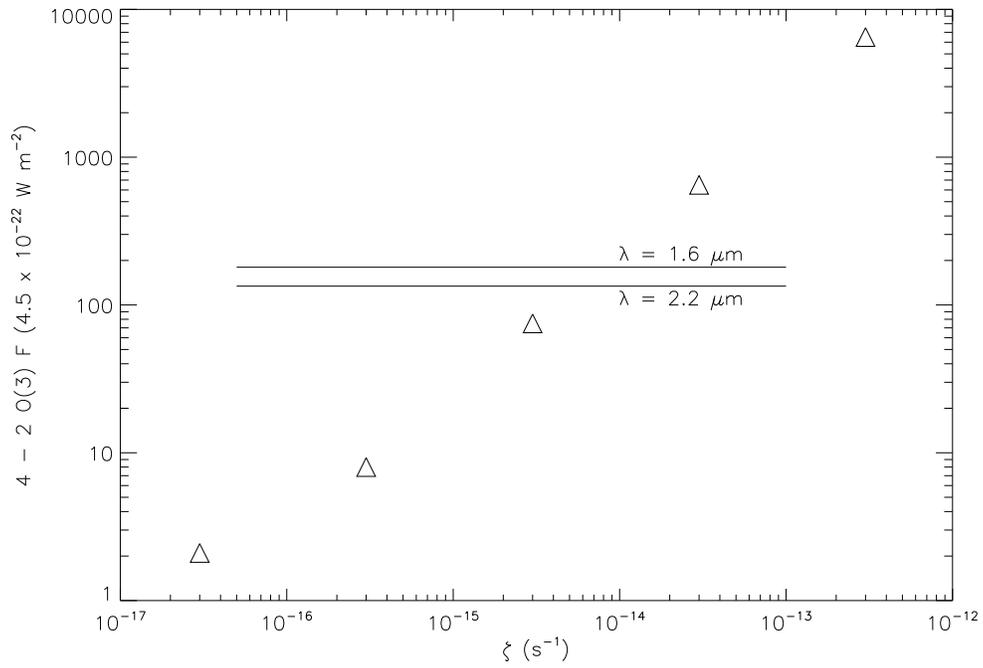}
\caption{Flux at Earth of the $4-2$ O(3) transition as a function of H$_2$ 
ionization rate for the RCM using model $(i)$. Solid horizontal lines represent 
UKIRT sensitivity in the H~(1.6~$\mu$m) and K~(2.2~$\mu$m) bands.} 
\label{fig8}
\end{figure}

In Fig. \ref{fig8}, we report the intensity of the line $4-2$~O(3) as a 
function of the H$_2$ ionization rate in the RCM. It is evident that the 
transition $4-2$~O(3) might be observed if the H$_2$ ionization rate is $\zeta
\ga 1 \times 10^{-14}$~s$^{-1}$, in agreement with \citet{LB95}. In this 
case, since the line intensity scales approximately with the ionization rate,
the line would be largely dominated by formation pumping rather than radiative
pumping. The crucial point is that XDRs may mantain large 
ionization rates up to hydrogen column densities of $N_{\rm H}\sim 10^{23} - 
10^{24}$~cm$^{-2}$ for X\textendash ray energies larger than 1~keV 
\citep{CP09}. 

In conclusion, we model H$_2$ formation pumping in standard dense clouds, 
taking into account the H/H$_2$ transition zone. The model, which includes 
recent laboratory data on H$_2$ formation, as well as the effects of the 
interstellar UV field, predicts the populations of gas\textendash phase H$_2$ 
molecules and their IR emission spectra. Calculations suggest that some 
vibrationally excited states of H$_2$ might be detectable towards lines of 
sight where significant destruction of H$_2$ occurs, such as X-ray sources.
These results also provide a possible explanation of the lack of detection to 
date of H$_2$ formation pumping in dark clouds.

\acknowledgments
CCP, SV and SC would like to thank the Royal Society for funding an exchange 
programme between UCL and Cagliari Observatory. We also would like to thank the
anonymous referee for hepful comments and suggestions that improved the
clarity of the paper.

\clearpage

\begin{table}
\caption{Normalized population distribution}
\begin{tabular} {cccccccc}
\hline
\hline
$v$ / $J$  & 0      & 1      & 2      & 3      & 4      & 5      & 6      \\
\hline
0          & \phantom{\tablenotemark{(a)}}0.0052\tablenotemark{(a)} & 0.0279 &
0.0048 & 0.0061 & 0.0006 &        &        \\

1          & 0.0052 & 0.0279 & 0.0048 & 0.0061 & 0.0006 &        &        \\
2          & 0.0058 & 0.0777 & 0.0116 & 0.0180 & 0.0016 &        &        \\
3          & 0.0213 & 0.1245 & 0.0245 & 0.0402 & 0.0073 & 0.0056 & 0.0005 \\
4          & 0.0362 & 0.1907 & 0.0539 & 0.0750 & 0.0108 & 0.0124 & 0.0020 \\
5          & 0.0135 & 0.0568 & 0.0266 & 0.0293 & 0.0052 & 0.0035 & 0.0011 \\
6          & 0.0026 & 0.0163 & 0.0057 & 0.0045 & 0.0009 &        &   \\
7          & 0.0021 & 0.0118 & 0.0039 & 0.0070 &        &        &   \\

\hline
\hline
\end{tabular}
\tablenotetext{(a)}{rotational populations of $v = 0$ are taken equal to the 
populations of $v = 1$ (see text).}

\end{table}

\begin{table}
\caption{Wavelength\textendash integrated emission from selected cloud models}
{\renewcommand{\arraystretch}{1.2}
\begin{tabular} {cccccccccc}
\hline
\hline
cloud model& $A_V$ & $\chi$ & $n_{\rm H}$  & \multicolumn{6}{c}{intensity} \\
& (mag) & & (cm$^{-3}$) &  \multicolumn{6}{c}{($10^{-6}$~ergs~s$^{-1}$
cm$^{-2}$~sr$^{-1}$)} \\
\cline{5-10}
& & & &  $(i)$ & $(ii)$ & $(iii)$
& $(iv-c)$ & $(v-c)$  & $(vi)$ \\ \hline
\multicolumn{10}{c}{dark clouds} \\
\phantom{\tablenotemark{(a)}}1\tablenotemark{(a)}

   & 100 & 1    & $10^6$ & 15.4  & 16.2  & 17.4  & 15.8  & 15.2  & 13.0  \\
2  & 100 & 10   & $10^6$ & 268   & 272   & 282   & 270   & 264   & 250   \\
3  & 100 & 100  & $10^6$ & 3328  & 3368  & 3472  & 3322  & 3290  & 3110  \\
4  & 100 & 1000 & $10^6$ & 49090 & 48714 & 49998 & 48288 & 48278 & 47002 \\
\multicolumn{10}{c}{translucent clouds} \\
5  & 5   & 1    & $10^3$ & 36.6  & 40.2  & 45.8  & 38.6  & 35.8  & 25.2  \\
6  & 10  & 1    & $10^3$ & 37.1  & 43.2  & 53.2  & 40.0  & 35.3  & 16.9  \\
7  & 5   & 1    & $10^4$ & 23.8  & 24.6  & 25.4  & 24.4  & 23.8  & 22.0  \\
8  & 10  & 1    & $10^4$ & 15.8  & 16.4  & 17.0  & 16.2  & 15.8  & 14.4  \\
9  & 5   & 1    & $10^5$ & 21.4  & 21.8  & 22.8  & 21.8  & 21.2  & 19.6  \\
10 & 10  & 1    & $10^5$ & 14.2  & 14.6  & 15.2  & 14.4  & 14.2  & 13.0  \\
11 & 5   & 1    & $10^6$ & 21.4  & 21.8  & 22.6  & 21.6  & 21.2  & 19.6  \\
12 & 10  & 1    & $10^6$ & 14.2  & 14.6  & 15.2  & 14.4  & 14.2  & 13.0  \\
\hline
\hline
\end{tabular}}

\tablenotetext{(a)}{Reference cloud model}
\end{table}

\begin{table}
\caption{Strongest emission lines}
\begin{tabular}{cccccccc}
\hline
\hline
wavelength &  transition & \multicolumn{5}{c}{residual 
intensity\tablenotemark{(a)}} & 
intensity \\ 
($\mu$m) & & \multicolumn{6}{c}{\phantom{AAAA}
($10^{-4}$~ergs~s$^{-1}$~cm$^{-2}$
sr$^{-1}$~$\mu$m$^{-1}$)} \\ 
\cline{3-8}
& & $(i)$ & $(ii)$ & $(iii)$
& $(iv-c)$ & $(v-c)$  & $(vi)$ \\ \hline

   0.90274 & $(4-1)$ Q(1)  & 1.5 &     &     &     &     & 2.4  \\
   1.06360 & $(2-0)$ S(7)  &     & 1.8 & 1.8 & 3.1 & 2.0 &      \\        
   1.11988 & $(3-1)$ S(9)  &     & 2.0 & 2.3 & 4.3 & 1.2 &      \\   
   1.18508 & $(3-1)$ S(3)  &     &     & 1.3 &     &     & 1.4  \\   
   1.23235 & $(3-1)$ S(1)  & 2.7 &     & 1.7 &     &     & 6.0  \\ 
   1.31342 & $(3-1)$ Q(1)  & 3.3 &     & 1.1 &     &     & 6.8  \\ 
   1.33475 & $(2-0)$ O(3)  & 4.1 &     & 1.3 &     &     & 8.2  \\ 
   1.40617 & $(2-0)$ Q(13) &     &     &     & 1.1 &     &      \\ 
   1.50913 & $(4-2)$ O(3)  & 7.4 &     & 2.0 &     &     & 12.4 \\ 
   1.71389 & $(1-0)$ S(8)  &     & 1.7 & 1.1 & 2.8 & 1.2 &      \\   
   1.74707 & $(1-0)$ S(7)  &     & 6.5 & 4.5 & 10.7& 6.1 &      \\
   1.83479 & $(1-0)$ S(5)  &     & 3.8 & 3.8 & 4.6 & 4.8 & 0.6  \\
   1.94392 & $(2-1)$ S(5)  &     & 1.1 & 1.6 &     & 1.6 & 0.3  \\
   2.15318 & $(2-1)$ S(2)  & 1.2 &     & 1.0 &     &     & 10.0 \\
   2.40524 & $(1-0)$ Q(1)  & 4.7 &     & 1.9 &     &     & 11.2 \\
   2.54981 & $(2-1)$ Q(1)  & 2.3 &     &     &     &     & 4.6  \\
   2.78490 & $(2-1)$ O(2)  & 1.3 &     &     &     &     & 14.9 \\
   2.80227 & $(2-1)$ Q(11) &     &     &     & 1.3 &     &      \\ 
   2.97273 & $(2-1)$ O(3)  & 6.2 &     & 2.0 &     &     & 12.6 \\
   3.54591 & $(0-0)$ S(16) &     & 1.2 &     &     &     &      \\
   3.72256 & $(0-0)$ S(14) &     & 2.2 &     &     &     &      \\
   3.83849 & $(1-1)$ S(15) &     & 1.2 &     &     &     &      \\
   3.84413 & $(0-0)$ S(13) &     & 1.2 &     &     &     &      \\
   4.07916 & $(2-2)$ S(15) &     & 1.3 &     &     &     &      \\
   4.17881 & $(0-0)$ S(11) &     & 2.4 &     & 2.1 & 1.3 &      \\
   4.22152 & $(1-1)$ S(12) &     & 1.2 &     & 1.2 &     &      \\
   4.41439 & $(1-1)$ S(11) &     & 2.2 & 1.1 & 3.0 & 1.6 &      \\
   4.95163 & $(1-1)$ S(9)  &     & 3.3 & 1.9 & 5.6 & 2.6 &      \\
\hline \hline
\end{tabular}
\tablenotetext{(a)}{peak line intensities larger than 
$1 \times 10^{-4}$~ergs~s$^{-1}$~cm$^{-2}$~sr$^{-1}$~$\mu$m$^{-1}$.} 
\end{table}

\begin{table}
\caption{Peak line emissions for the cloud models reported in Table~2}
\begin{tabular}{cccccccccc}
\hline
\hline
cloud model\tablenotemark{(a)} & \multicolumn{9}{c}{intensity} \\ 
& \multicolumn{9}{c}
{($10^{-4}$~ergs~s$^{-1}$~cm$^{-2}$ sr$^{-1}$~$\mu$m$^{-1}$)} \\
\cline{2-10}
& \multicolumn{3}{c}{$(i)$}  
& \multicolumn{3}{c}{$(ii)$}  
& \multicolumn{3}{c}{$(iii)$}  \\ 
& L1\tablenotemark{(b)} & L2 & L3
& L1 & L2 & L3
& L1 & L2 & L3 \\
\hline

1  & 19.9 & 0.05 & 16.0  & 12.7 & 6.5  & 11.7  & 14.4 & 4.6  & 13.1  \\
2  & 274  & 17.1 & 354   & 227  & 54.9 & 327   & 239  & 42.9 & 336   \\
3  & 3407 & 204  & 4082  & 2880 & 537  & 3777  & 3024 & 420  & 3886  \\
4  & 19977& 32133& 124859& 15834& 33556& 123270& 17034& 32984& 123902\\
5  & 54.6 & 0.19 & 38.8  & 23.9 & 30.2 & 22.2  & 30.4 & 21.1 & 27.3  \\
6  & 71.6 & 0.17 & 46.0  & 17.0 & 53.4 & 16.6  & 28.4 & 37.2 & 25.6  \\
7  & 24.4 & 0.11 & 19.8  & 19.3 & 5.1  & 17.1  & 20.4 & 3.6  & 17.9  \\
8  & 16.4 & 0.07 & 13.2  & 12.8 & 3.6  & 11.3  & 13.6 & 2.5  & 11.9  \\
9  & 22.0 & 0.08 & 17.7  & 17.2 & 4.7  & 15.1  & 18.1 & 3.3  & 15.9  \\
10 & 14.8 & 0.05 & 11.9  & 11.4 & 3.4  & 10.0  & 12.1 & 2.4  & 10.6  \\
11 & 24.0 & 0.07 & 20.2  & 18.9 & 4.6  & 17.2  & 20.2 & 3.2  & 18.2  \\
12 & 16.2 & 0.04 & 13.7  & 12.6 & 3.3  & 11.6  & 13.5 & 2.3  & 12.2  \\

& & & & & & & & & \\

& \multicolumn{3}{c}{$(iv-c)$}   
& \multicolumn{3}{c}{$(v-c)$}  
& \multicolumn{3}{c}{$(vi)$} \\
& L1 & L2 & L3
& L1 & L2 & L3
& L1 & L2 & L3 \\
\hline

1  & 12.5 & 10.7 & 11.2  & 12.7 & 6.1  & 12.1  & 12.5 & 0.04 & 11.3  \\
2  & 225  & 79.2 & 324   & 226  & 52.5 & 329   & 225  & 16.4 & 322   \\
3  & 2855 & 752  & 3742  & 2877 & 520  & 3816  & 2864 & 193  & 3725  \\
4  & 15779& 34549& 123085& 15975& 33603& 123764& 15968& 31752& 122916\\
5  & 23.1 & 49.9 & 20.8  & 23.7 & 28.3 & 23.3  & 22.9 & 0.14 & 20.5  \\       
6  & 15.6 & 88.2 & 14.0  & 16.7 & 49.9 & 18.6  & 15.5 & 0.09 & 13.8  \\
7  & 19.2 & 8.4  & 16.9  & 19.3 & 4.8  & 17.3  & 19.1 & 0.10 & 16.8  \\
8  & 12.7 & 5.9  & 11.2  & 12.8 & 3.4  & 11.5  & 12.6 & 0.06 & 11.1  \\
9  & 17.0 & 7.8  & 14.8  & 17.1 & 4.4  & 15.2  & 17.0 & 0.07 & 14.8  \\
10 & 11.3 & 5.5  & 9.8   & 11.4 & 3.2  & 10.1  & 11.3 & 0.05 & 9.8   \\
11 & 18.8 & 7.6  & 16.9  & 18.9 & 4.3  & 17.4  & 18.8 & 0.06 & 16.9  \\
12 & 12.5 & 5.5  & 11.3  & 12.6 & 3.1  & 11.7  & 12.5 & 0.04 & 11.3  \\

\hline \hline
\end{tabular}

\tablenotetext{(a)}{Numbers refer to Table~2}

\tablenotetext{(b)}{L1: $(4-2)$ O(3), L2: $(1-0)$ S(7), L3: $(1-0)$ Q(1)}
\end{table}

\begin{table}
\caption{Entry and exit rates of the upper levels of the transitions
$(4-2)$ O(3) and $(1-0)$~S(7) computed for the RCM}
\hspace{-1.75cm}
\begin{tabular}{cccccccccc}
\hline
\hline
$A_V$ & \multicolumn{7}{c}{$(v,J)$} & \multicolumn{2}{c}{rates (s$^{-1}$)} \\
(mag) & \multicolumn{2}{c}{$(4,1)$} & \multicolumn{5}{c}{$(1,9)$} & & \\
& $(i)$ & & $(ii)$ & $(iii)$ & $(iv-c)$ & $(v-c)$ & & & \\
& entry & exit & entry & entry & entry & entry & exit & & \\
0.1 & 1.3(-16) & 3.0(-17) & 2.6(-19) & 1.9(-18) & 1.2(-17) & 2.0(-17) & 
3.0(-17) & ${\cal R} n_1 \delta_i$ & $\zeta$ \\
    & 6.1(-16) & 3.1(-06) & 1.7(-16) & 1.3(-16) & 3.0(-16) & 1.4(-16) & 
7.1(-07) & $\sum_{j > i} A_{ji} n_j/n_2$ & $\sum_{i > j} A_{ij}$ \\ 
    & 3.8(-16) & 1.5(-09) & 1.5(-19) & 1.0(-19) & 1.9(-19) & 1.2(-19) & 
1.2(-09) & $\sum_{j \ne i} W_{ji} n_j/n_2$ & $\sum_{i \ne j} W_{ij}$ \\ 
    & 1.2(-16) & 2.0(-15) & 9.9(-18) & 7.1(-18) & 1.7(-17) & 9.5(-18) 
& 3.7(-08) & $\sum_{j \ne i} C_{ji} n_j/n_2$ & $\sum_{i \ne j} C_{ij}$ \\

\multicolumn{10}{c}{ } \\

0.5 & 6.4(-19) & 3.0(-17) & 1.1(-19) & 1.1(-19) & 6.4(-19) & 1.1(-18) &
3.0(-17) & & \\
    & 2.6(-17) & 3.1(-06) & 9.5(-18) & 7.3(-18) & 1.7(-17) & 7.8(-18) & 
7.1(-07) & & \\
    & 1.6(-17) & 2.5(-10) & 9.5(-22) & 6.9(-22) & 1.2(-21) & 7.5(-22) & 
1.5(-10) & & \\
    & 5.3(-18) & 1.1(-16) & 3.8(-20) & 2.6(-20) & 1.2(-19) & 3.5(-20) & 
2.9(-09) & & \\

\multicolumn{10}{c}{ } \\

1.0 & 3.0(-18) & 3.0(-17) & 2.2(-17) & 6.7(-20) & 2.7(-19) & 4.6(-19) & 
3.0(-17) & & \\
    & 1.9(-18) & 3.1(-06) & 3.9(-18) & 3.0(-18) & 6.9(-18) & 3.2(-18) & 
7.1(-07) & & \\ 
    & 7.4(-19) & 3.5(-11) & 3.3(-23) & 2.8)-23) & 4.3(-23) & 2.5(-23) & 
1.5(-11) & & \\ 
    & 7.1(-19) & 4.3(-17) & 8.2(-21) & 5.3(-21) & 1.3(-20) & 7.3(-21) & 
1.7(-09) & & \\

\multicolumn{10}{c}{ } \\

2.5 & 3.0(-18) & 3.0(-17) & 5.6(-21) & 3.4(-20) & 2.5(-19) & 4.3(-19) & 
3.0(-17) & & \\  
    & 2.5(-22) & 3.1(-06) & 3.7(-18) & 2.8(-18) & 6.5(-18) & 3.0(-18) & 
7.1(-07) & & \\ 
    & 2.8(-22) & 1.6(-13) & 5.5(-26) & 9.6(-26) & 5.6(-26) & 3.1(-26) & 
2.4(-14) & & \\
    & 4.7(-19) & 4.7(-17) & 3.3(-21) & 5.2(-21) & 1.2(-20) & 6.5(-21) & 
1.6(-09) & & \\

\multicolumn{10}{c}{ } \\

5.0 & 3.0(-18) & 3.0(-17) & 5.6(-21) & 3.4(-20) & 2.5(-19) & 4.3(-19) & 
3.0(-17) & & \\
    & 3.5(-19) & 3.1(-06) & 3.7(-18) & 2.8(-18) & 6.5(-18) & 3.0(-18) &
7.1(-07) & &  \\ 
    & 2.3(-27) & 3.9(-17) & 1.5(-29) & 5.2(-29) & 4.3(-30) & 2.2(-30) & 
1.0(-18) & & \\ 
    & 4.7(-19) & 4.7(-17) & 3.3(-21) & 5.2(-21) & 1.2(-20) & 6.5(-21) & 
1.6(-09) & & \\

\multicolumn{10}{c}{ } \\

10. & 3.0(-18) & 3.0(-17) & 5.6(-21) & 3.4(-20) & 2.5(-19) & 4.3(-19) & 
3.0(-17) & & \\
    & 3.5(-19) & 3.1(-06) & 3.7(-18) & 5.2(-29) & 6.5(-18) & 3.0(-18) & 
7.1(-07) & & \\ 
    &          & 4.6(-24) &          & & & & 4.6(-27) & & \\
    & 4.5(-19) & 3.9(-17) & 3.3(-21) & 5.2(-21) & 1.2(-20) & 6.5(-21) & 
1.6(-09) & & \\

\multicolumn{10}{c}{ } \\

50. & 3.0(-18) & 3.0(-17) & 5.6(-21) & 3.4(-20) & 2.5(-19) & 4.3(-19) & 
3.0(-17) & & \\
    & 3.5(-19) & 3.1(-06) & 3.7(-18) & 5.2(-29) & 6.5(-18) & 3.0(-18) & 
7.1(-07) & & \\ 
    & & & & & & & & &\\
    & 4.5(-19) & 3.9(-17) & 3.3(-21) & 5.2(-21) & 1.2(-20) & 6.5(-21) & 
1.6(-09) & & \\

\hline
\hline
\end{tabular}
\end{table}

\end{document}